\documentclass[10pt]{article}
\usepackage{graphicx}
\usepackage[ansinew]{inputenc}
\usepackage{amsmath, amssymb, amsfonts, amsthm, bm, natbib} % write standard N, R, Z, Q symbols for number sets.
\usepackage{dsfont}
\usepackage{hyperref}
\graphicspath{{/home/vinicius/Dropbox/Papers/JAGS_MEP/JSS/Pictures/}} % Graphics Path to find your pictures
\newcommand{\sex}{\mbox{\tiny sex}\,}
\newcommand{\age}{\mbox{\tiny age}\,}

\usepackage{listings}
\usepackage[usenames,dvipsnames]{color}    
\begin{document} %%%%%%%
%%%%%%%%%%%%%%%%%%%%%%%%

% Title
\begin{flushleft}
{\Large \bf pexm: a JAGS module for applications involving the piecewise exponential distribution.}
\end{flushleft}

\vspace{5pt}

{\flushleft 
{\bf Vin\'{i}cius D. Mayrink, \, Jo\~{a}o D. N. Duarte \, and \, F\'{a}bio N. Demarqui.} \\
{\small \em Departamento de Estat\'{i}stica, ICEx, Universidade Federal de Minas Gerais.}}

\vspace{10pt}

\begin{abstract}
In this study, we present a new module built for users interested in a programming language similar to \texttt{BUGS} to fit a Bayesian model based on the piecewise exponential (PE) distribution. The module is an extension to the open-source program \texttt{JAGS} by which a Gibbs sampler can be applied without requiring the derivation of complete conditionals and the subsequent implementation of strategies to draw samples from unknown distributions. The PE distribution is widely used in the fields of survival analysis and reliability. Currently, it can only be implemented in \texttt{JAGS} through methods to indirectly specify the likelihood based on the Poisson or Bernoulli probabilities. Our module provides a more straightforward implementation and is thus more attractive to the researchers aiming to spend more time exploring the results from the Bayesian inference rather than implementing the Markov Chain Monte Carlo (MCMC) algorithm. For those interested in extending \texttt{JAGS}, this work can be seen as a tutorial including important information not well investigated or organized in other materials. Here, we describe how to use the module taking advantage of the interface between \texttt{R} and \texttt{JAGS}. A short simulation study is developed to ensure that the module behaves well and a real illustration, involving two PE models, exhibits a context where the module can be used in practice. 
\vspace{3pt}
{\flushleft keywords: MCMC, Bayesian Inference, Semiparametric, Survival Analysis.}
\end{abstract}

\begin{figure}[b]
{\noindent \small \emph{Address for correspondence:}
Vin\'{i}cius Mayrink, Departamento de Estat\'{i}stica, ICEx, UFMG. Av. Ant\^{o}nio Carlos, 6627, Belo Horizonte, MG, Brazil, 31270-901. E-mail: vdm@est.ufmg.br}
\end{figure}

%%%%%%%%%%%%%%%%%%%%%%%%%%%%%%%%%%%%%%%%%%%%%%%%%%%%%%%%%%%
\section[Introduction]{Introduction} \label{secintro}
%%%%%%%%%%%%%%%%%%%%%%%%%%%%%%%%%%%%%%%%%%%%%%%%%%%%%%%%%%%

The piecewise exponential (PE) modeling is a simple approach extensively used in survival analysis and reliability to approximate the distribution of event-time data; possibly investigating the association of the time response with explanatory variables. Applications related to clinical data can be easily found in the literature; for example: leukemia \citep{Bres74}, gastric cancer \citep{Gam91}, kidney infection \citep{Sahu97,Ibra01}, breast cancer \citep{Sinha99}, melanoma \citep{Dem14} and hospital mortality \citep{Cla02}. In the context of reliability, consider the maintenance analysis in \cite{Rig00} and the telecommunication study in \cite{Kim91} and \cite{Gam94}, among many other references.

The PE models are classified as semiparametric due to their flexibility to represent the hazard function without making strong and restrictive parametric assumptions regarding the shape of this component; this point makes the PE distribution very attractive for survival analysis. Its semiparametric nature allows considerable generality and enough structure for useful interpretations in any application. The popularity of semiparametric methods for univariate survival data started with \cite{Cox72} on the proportional hazards model. \cite{Bres72} and \cite{Bres74} proposed the use of the PE distribution to replace the baseline term and \cite{Kalb73} investigated the grid configuration. In this paper, we are focused on PE models under the Bayesian approach for inference, which has been widely considered in the literature. See \cite{Ibra01} for a broad presentation of Bayesian survival models and \cite{Sinha97} for a review on Bayesian semiparametric frameworks (including the PE) based on either the hazard or the intensity function.

Implementing a PE model is an obstacle for those interested in a Bayesian analysis and having limited programming skills. In many cases, the main source of this difficulty is the use of an indirect method to draw samples from the unknown joint posterior distribution. The Markov Chain Monte Carlo (MCMC) algorithm Gibbs sampling \citep[see][]{Gem84,Gelf90,Gam06} is widely applied for this task, and its implementation can be very demanding due to the sequential simulations from complete conditional posterior distributions. This is particularly true when the conjugate analysis is not possible leading to a Gibbs sampler whose steps depend on other methods -- e.g., Metropolis-Hastings \citep{Met53,Has70}, Adaptive Rejection Sampling \citep{Gil92, Gil92c}, Adaptive Rejection Metropolis Sampling \citep{Gil95} and Slice Sampling \citep{Neal03}.

As an alternative to simplify the Gibbs sampler implementation for a given model, one can take advantage of programs based on a dialect of the \texttt{BUGS} language \citep{Tho92, Lun12} using the mathematical formalism of directed acyclic graphs to define joint densities. These programs build the whole structure of complete conditionals and their corresponding sampling strategies requiring from the user only the essential information related to the model fit (priors, likelihood and the MCMC setup). In this category, \texttt{WinBUGS} \citep{Spie03}, \texttt{OpenBUGS} \citep{Spie14} and \texttt{JAGS} \citep{Plu03} are three well known options. In particular, \texttt{JAGS} is an appealing case since it is open-source (released under a free copyleft license) and extensible allowing users with \texttt{C++} knowledge to create functions, monitors, distributions and samplers for a new module. These features enable the potential formation of a future \texttt{JAGS} contributor community similar to that of the software \texttt{R} \citep{R}.

Other alternatives are available as a support for the Bayesian computing. A well known case is the platform \texttt{Stan} (\url{mc-stan.org}), having an interface with \texttt{R} through the package \texttt{rstan} \citep{Rstan}. The MCMC method applied here is called No-U-Turn Sampling \citep{Hoff14}, being an extension of the Hamiltonian Monte Carlo method. Implementing new distributions in \texttt{Stan} is straightforward and it does not require the installation of new modules. However, due to the Hamiltonian dynamics to improve sampling, one cannot set discrete prior distributions. This is perhaps the greatest weakness of \texttt{Stan}. Another possibility to deal with Bayesian modeling in \texttt{R} is the package \texttt{BayesX} \citep{Brez05}. However, this alternative was designed for the framework of generalized linear models, where the distribution of the response variable belongs to the exponential family. This is not the case for survival models assuming the PE distribution. Bayesian computing can also be developed via \texttt{INLA} \citep{Rue09}; further details in \url{r-inla.org}. This option performs approximate Bayesian inference on the class of latent Gaussian models. It handles non-Gaussian likelihoods, but the presence of non-Gaussian latent components can be a limitation; for example, assuming a discrete or a multimodal distribution for a random effect would be problematic.  

The main contribution of this work is to present a new \texttt{JAGS} module (\texttt{pexm}) for Bayesian analyses involving the PE distribution. An \texttt{R} package, having the same name, was developed to install and load the tool. The proposed module was created to fill a gap existing in \texttt{JAGS} related to the absence of the PE distribution. The \texttt{pexm} allows a friendlier and cleaner model implementation, being attractive for many users. Computational speed is also an advantage of the proposal with respect to the usual implementation based on the Poisson zeros-trick. Another important contribution is a description to clarify some aspects related to the topic ``extending \texttt{JAGS}'', which is not well documented in the literature. There are no formal tutorials exploring the technical elements detailed in Appendix A. The \texttt{JAGS} manuals \citep{Plu10,Plu17} do not currently contain the mentioned information, however, these materials can be updated in the future.  

The outline of this paper is as follows: in Section 2, we describe the elements of the PE distribution implemented in the proposed \texttt{JAGS} module. In Section 3, we present an overview on how the module was created, explain how to use it from within \texttt{R}, discuss the main simplifications with respect to an implementation without the module and develop a simulation study to verify the performance. Section 4 shows results (with and without the module) in a real application involving two frailty models to fit a kidney infection data \citep{Ibra01}; the rates and log-rates in adjacent intervals are correlated via gamma and normal priors, respectively. Finally, Section 5 presents the conclusions and final remarks.

%%%%%%%%%%%%%%%%%%%%%%%%%%%%%%%%%%%%%%%%%%%%%%%%%%%%%%%%%%%
\section[The piecewise exponential distribution]{The piecewise exponential distribution} \label{secped}
%%%%%%%%%%%%%%%%%%%%%%%%%%%%%%%%%%%%%%%%%%%%%%%%%%%%%%%%%%%

The PE distribution requires the specification of a grid $\tau = \{ a_1, a_2, \ldots, a_m \}$ partitioning the time axis into $m$ intervals. Let $a_1 = 0$, $a_m < +\infty$ and $a_{m+1} = +\infty$; the latter is defined only for notation purposes and it is not expressed in $\tau$. Conditional on this time grid, the failure rate $\lambda_j$ is assumed constant inside the corresponding interval: $I_j = (a_{j},a_{j+1}]$ for $j = 1,\ldots,m-1$ and $I_m = (a_{m}, +\infty)$ for $j = m$. This step function formulation provides a discrete version of the true unknown hazard function in a survival model. Assume that $T$ is a continuous non-negative random variable representing the survival times of subjects in a population. Let $f(t)$ denote the probability density function (pdf) of $T$, $F(t)$ is the associated cumulative distribution function (cdf), $S(t) = 1-F(t)$ is the survival function, $h(t)$ is the hazard function and, finally, $H(t)$ represents the cumulative hazard function. If $T \sim \mbox{PE}(\lambda,\tau)$ with $\lambda = \{ \lambda_1, \ldots, \lambda_m \}$, then:
\begin{equation}
\label{eqht}
h(t) = \lambda_j, \;\; \mbox{if} \;\; t \in I_j, \;\; j = 1, \ldots, m.
\end{equation}

Naturally, the quality of the approximation to the real hazard function depends on a suitable choice of the time grid $\tau$. In \cite{Bres72} the author suggests the use of the observed failure times as the limits of the intervals $I_j$; however, depending on the sample size, this can lead to a non-parsimonious model with many rates $\lambda_j$ to be estimated. In contrast, \cite{Kalb73} indicates that the grid should be selected independently of the data. The study explores different interval sizes configuring a regular grid. When choosing $\tau$ without using the data, one might consider two aspects: small $m$ leads to a poor discretization and a large $m$ determines a non-parsimonious model. Irregular grids are also allowed in the analysis and this could be a reasonable strategy in applications assuming higher frequency of failures in certain parts of the time axis; i.e., wider intervals would be specified for time regions where few events are expected to be observed.

The cumulative hazard function is expressed as follows:
\begin{equation}
\label{eqHt}
H(t) = \left\{ \begin{array}{l}
\lambda_1 \, t, \;\; \mbox{if} \;\; t \in I_1;\\
\lambda_j (t-a_{j}) + \sum_{i = 1}^{j-1} \lambda_i (a_{i+1}-a_{i}), \;\; \mbox{if} \;\; t \in I_j, \; j = 2, \ldots, m. 
\end{array}  \right.
\end{equation}
We can use (\ref{eqHt}) to evaluate the survival function at a time point $t$ through the well known equality $S(t) = \exp\{ -H(t) \}$. The pdf of $T$ can be easily identified by simply taking the derivative of $F(t) = 1 - S(t)$ with respect to $t$. This provides $f(t) = h(t) \exp\{ -H(t) \}$, where $h(t)$ is given in (\ref{eqht}). 

The quantile is an information of interest in many data analyses. In the context of the PE distribution, this quantity can be calculated via the relationship $H(t) = -\ln(1-F(t))$. Given a probability $p$, the corresponding quantile is the value $t^*$ such that $H(t^*) = -\ln(1-p)$. Let $w(p) = -\ln(1-p)$ and recall that $H(a_{m+1}) = +\infty$. Then, the calculation of $t^*$ can be expressed by:
\begin{equation}
\label{eqtstar}
t^* = \left\{ \begin{array}{l}
a_1 + \dfrac{w(p)}{\lambda_1}, \;\; \mbox{if} \;\; w(p) \leq H(a_2) = \lambda_1; \vspace{5pt} \\
a_j + \dfrac{w(p) -  \sum_{i=1}^{j-1} \lambda_i (a_{i+1}-a_{i})}{\lambda_j}, \;\; \mbox{if} \;\; H(a_{j}) < w(p) \leq H(a_{j+1}), \, j = 2, \ldots, m.
\end{array}  \right.
\end{equation}
Note that we can easily compute the median of the PE distribution by setting $p = 0.5$ in (\ref{eqtstar}). Those readers interested in details about the first two moments of the PE distribution can refer to \cite{Bar96}.  

%%%%%%%%%%%%%%%%%%%%%%%%%%%%%%%%%%%%%%%%%%%%%%%%%%%%%%%%%%%
\section[The JAGS module]{The \texttt{JAGS} module.} \label{secmodule}
%%%%%%%%%%%%%%%%%%%%%%%%%%%%%%%%%%%%%%%%%%%%%%%%%%%%%%%%%%%

\texttt{JAGS} \citep[``Just Another Gibbs Sampler''; see][]{Plu03} is a general-purpose, open-source, cross-platform graphical modeling program using a set of MCMC methods for stochastic simulations to generate samples from the joint posterior distribution of the parameters in a Bayesian model. These samples are further used for inferences regarding the target parameters and the model fit. A recent version of \texttt{JAGS} can be downloaded from \url{http://mcmc-jags.sourceforge.net} and it is also available as a package for different Linux distributions.

A special feature of \texttt{JAGS} is that it can be modular and extensible with new functionalities, which can be loaded whenever required. A module can be created to accommodate all types of probability distributions (continuous or discrete, univariate or multivariate). In order to write these extensions, the knowledge of \texttt{C++} and object oriented programming is necessary; however, we emphasize that installing the extensions can be extremely automatic depending on the operational system.       

Technical manuals providing a comprehensive coverage on how to build a \texttt{JAGS} module were not available in the literature at the time of writing this material. We follow the steps indicated in \cite{Wab14}, which can be seen as a short tutorial exploring the key aspects to add custom distributions to \texttt{JAGS}. Those aspects not covered in this reference, but important to implement the PE distribution, are given careful attention in Appendix A. For instance, the authors in \cite{Wab14} work with two examples: the Bernoulli distribution for didactic purposes and the Wiener diffusion first-passage time distribution. In both cases the input values are scalar quantities; i.e., the paper does not explore distributions with vectors as arguments, which is the case for the $\mbox{PE}(\lambda,\tau)$. This issue can be easily addressed by simply modifying the parent class type expressed in the code from scalar to vector (see Appendix A).

The PE distribution implementation described in Appendix A is in fact a more complete example displaying the use of important resources available to create a \texttt{JAGS} module. Besides implementing the log-density and the sample generating process allowing the MCMC simulations in a Gibbs sampler, we have built auxiliary functions returning the pdf, cdf, quantile, hazard and cumulative hazard of the $\mbox{PE}(\lambda, \tau)$ at a given time $t$ or probability $p$. In addition, the module includes a routine to check whether the values in $\lambda$ and $\tau$ are consistent with the specifications shown in Section \ref{secped}. Hence, Appendix A contains essential complementary information \citep[with respect to][] {Wab14} to those readers interested in extending \texttt{JAGS}.

%%%%%%%%%%%%%%%%%%%%%%%%%%%%%%%%%%%%%%%%%%%%%%%%%%%%%%%%%%%
\subsection[Using the module in the R environment]{Using the module in the \texttt{R} environment.} \label{secR}
%%%%%%%%%%%%%%%%%%%%%%%%%%%%%%%%%%%%%%%%%%%%%%%%%%%%%%%%%%%

In this section, we discuss how to use the proposed module in association with \texttt{R}. This is achieved through the package \texttt{rjags} \citep{Plu16} providing an interface between \texttt{R} and \texttt{JAGS}. Installing the new module is straightforward as indicated in Appendix A. The procedure basically involves the installation of an \texttt{R} package called \texttt{pexm}. The package was designed for this task and also contains tools to help loading the module into the \texttt{R} environment. The \texttt{pexm} source can be downloaded from two repositories: \texttt{GitHub} and \texttt{Sourceforge} (see more details in Appendix \ref{secA3}). Hereafter, we assume that a recent version of \texttt{JAGS}, \texttt{R}, \texttt{rjags} and the proposed \texttt{pexm} module are correctly installed in the computer. A confirmation that the module is ready to be used in \texttt{R} can be obtained by simply typing the following commands in the console: 
\begin{lstlisting}
R> library("pexm")
R> loadpexm() 
\end{lstlisting}

In order to illustrate how to use the module, we have developed a simple example taking into account the \texttt{R} package \texttt{msm} \citep{Jack11} to generate the data. This package provides functions to fit continuous-time Markov and hidden Markov multi-state models and, in particular, includes the function \texttt{rpexp} to draw samples from the PE distribution. We can also use \texttt{dpexp}, \texttt{ppexp} and \texttt{qpexp} to explore the pdf, cdf and quantiles.

Let $T_i$ be the survival time of the $i$-th individual, $i = 1,\ldots,n$. In our illustration, we set the real rates $\lambda = (0.3, 0.6, 0.8, 1.3)^\top$, the time grid $\tau = \{ 0, 2, 3, 5\}$, $n = 1{,}000$ and assume $T_i \sim \mbox{PE}(\lambda, \tau)$ to generate the data. The following objects are loaded into the \texttt{R} workspace: \texttt{t} is the observed vector $(T_1,\ldots,T_n)^\top$, \texttt{n} is the length of \texttt{t}, \texttt{tau} is the grid vector, \texttt{m} is the length of \texttt{tau}, \texttt{pq} is a vector of probabilities for which the quantile should be evaluated, and \texttt{nq} is the length of \texttt{pq}. 

\vspace{5pt}
\noindent \textbf{Code chunk CC.1:} 
\vspace{-5pt}
\begin{lstlisting}[numbers=left] 
R> data <- list(t = t, n = n, tau = tau, m = m, pq = pq, nq = nq)
R> parameters <- c("lambda", "ht100", "Ht100", paste0("q[",1:nq,"]"), paste0("loglik[",1:n,"]"), paste0("St[",1:n,"]"))
R> inits1 <- list( lambda = c(0.1, 0.5, 1, 2), .RNG.name = "base::Super-Duper", .RNG.seed = 1 ) 
R> inits2 <- list( lambda = c(0.5, 1.0, 1.5, 2.5), .RNG.name = "base::Wichmann-Hill", .RNG.seed = 2 )
R> burnin <- 1000
R> lag <- 1
R> npost <- 2000
R> Mjags <- rjags::jags.model(file = "Model_pex.R", data = data, inits = list(inits1, inits2), n.chains = 2, n.adapt = burnin)
R> output <- rjags::coda.samples(Mjags, variable.names = parameters, n.iter = npost, thin = lag)
\end{lstlisting}

The Code chunk CC.1 shows a simple \texttt{R} script to organize the workspace information and then access \texttt{JAGS} for the MCMC simulation. Note that we specify: a list object (\texttt{data}) containing \texttt{t}, \texttt{n}, \texttt{tau}, \texttt{m}, \texttt{pq} and \texttt{nq}; a string vector (\texttt{parameter}) indicating which chains must be saved; two list objects (\texttt{inits1} and \texttt{inits2}) providing initial values for $\lambda$ and setting details about random seeds in \texttt{JAGS}. The MCMC setup is determined through the elements: size of the burn-in period (\texttt{burnin}), thinning interval of the chains (\texttt{lag}) and the number of posterior samples required for inference (\texttt{npost}). The last two commands, \texttt{jags.model} and \texttt{coda.samples}, are functions from the library \texttt{rjags} to compile the Bayesian model in \verb'Model_pex.R' (written in the \texttt{JAGS} syntax) and to run the corresponding MCMC algorithm, respectively. It is important to emphasize that the Code chunk CC.1 assumes that the \texttt{R} working directory is the same one where the model script \verb'Model_pex.R' is located. Note that we choose to handle the results in a format that can be conveniently analyzed via the \texttt{R} package \texttt{coda} \citep{Plu06}. Results are saved in the object \texttt{output}. The posterior estimates discussed ahead are based on the two chains generated for each parameter.

\vspace{5pt}
\noindent \textbf{Code chunk CC.2} 
\vspace{-5pt}
\begin{lstlisting}[numbers=left]
model{
  for (i in 1:n) {
    t[i] ~ dpex(lambda[], tau[])
    St[i] <- 1 - ppex(t[i], lambda[], tau[])
    loglik[i] <- log(dpex(t[i], lambda[], tau[]))
  }
  Ht100 <- hcpex(t[100], lambda[], tau[])
  ht100 <- hpex(t[100], lambda[], tau[])
  for (j in 1:m) { lambda[j] ~ dgamma(0.01, 0.01) }
  for (k in 1:nq) { q[k] <- qpex(pq[k], lambda[], tau[]) }  
}
\end{lstlisting}

The content of the \texttt{JAGS} model script \verb'Model_pex.R' is presented in the Code chunk CC.2. This code indicates the likelihood $T_i \sim \mbox{PE}(\lambda,\tau)$ and the prior specifications $\lambda_j \sim \mbox{Ga}(0.01,0.01)$ for $j = 1,\ldots,m$ (mean $1$ and variance $100$). Note that in each iteration of the MCMC, the survival function \texttt{St[i]} and the log-density \texttt{loglik[i]} are evaluated for each $t_i$ together with the quantile \texttt{q[k]} for the probabilities in \texttt{pq}. In addition, the cummulative hazard function \texttt{Ht100} and the hazard function \texttt{ht100}, related to the time point $t_{100}$, are monitored in this example. As it can be seen, the elements defined in \verb'Model_pex.R' explore all functionalities implemented for the PE distribution in the new module.

%%%%%%%%%%%%%%%%%%%%%%%%%%%%%%%%%%%%%%%%%%%%%%%%%%%%%%%%%%%
\subsection[Strategies to indirectly specify the likelihood]{Strategies to indirectly specify the likelihood.} \label{secPois}
%%%%%%%%%%%%%%%%%%%%%%%%%%%%%%%%%%%%%%%%%%%%%%%%%%%%%%%%%%%

Without the proposed \texttt{pexm} module, one can use either the Poisson or the Bernoulli distribution to indirectly specify the likelihood function. These strategies are commonly known as zeros-trick \citep{Lun12} and ones-trick \citep{Spie03}, respectively. Let $l_i = \ln[f(t_i|\lambda,\tau)]$, then the likelihood can be written as:
$$
  f(t_1,\ldots,t_n| \theta) \; = \; \prod_{i = 1}^n \exp\{ l_i \} \; = \;  \prod_{i = 1}^n \dfrac{\exp\{ -(-l_i) \} (-l_i)^0}{0!} \; = \;  \prod_{i = 1}^n P(X_i = 0),
$$
where $X_i \sim \mbox{Pois}(-l_i)$. In order to ensure a positive Poisson mean, we can select a constant $C > 0$ and use:  
$$
  \exp\{-C\} \; f(t_1,\ldots,t_n| \theta) \; = \; \prod_{i = 1}^n \dfrac{\exp\{ -(-l_i+C) \} (-l_i+C)^0}{0!} \; = \; \prod_{i = 1}^n P(X_i^* = 0),
$$
where $X_i^* \sim \mbox{Pois}(-l_i+C)$. We must choose $C$ such that $-l_i + C > 0$ for all $i = 1,\ldots,n$. Note that adding $C$ does not affect the inference results, since it is equivalent to multiplying the posterior distribution by a constant term.

\vspace{5pt}
\noindent \textbf{Code chunk CC.3} 
\vspace{-5pt}
\begin{lstlisting}[numbers=left]
model {
  C <- 10000
  for (i in 1:n) {
      zeros[i] ~ dpois( zero_mean[i] )
      zero_mean[i] <- -l[i] + C
      l[i] <- ... # log-likelihood here.
  }
  ... # Priors and other terms here.
}
\end{lstlisting}

This strategy can be applied in \texttt{JAGS} using the syntax shown in Code chunk CC.3. Note that the log-likelihood of the $\mbox{PE}(\lambda,\tau)$ for the $i$-th observation must be specified for the object \texttt{l[i]}. This task for the PE context is not simple compared to other distributions. A possible approach is to replace the vector \texttt{t} and \texttt{tau} by other objects accounting for the position of each observed time point in the grid. These objects must be created in the \texttt{R} environment and passed to \texttt{JAGS} as a data argument (see \texttt{data} in the Code chunk CC.1) together with \texttt{zero[i] = 0} for all $i$. They must allow the calculation of $H(t)$ in \texttt{JAGS}, which in turn can be used to obtain $\ln[f(t)] = \lambda_j - H(t)$ (if $t \in I_j$) and $S(t) = \exp\{-H(t)\}$. After the MCMC simulation, the quantiles can be calculated in \texttt{R} using the posterior samples of $\lambda$ and the function \texttt{qpexp} (package \texttt{msm}).

Again, we call attention to the fact that the Bernoulli distribution can also be used in a similar strategy expressing the model likelihood as $\prod_{i=1}^n P(X_i^* = 1)$ with $X_i^* \sim \mbox{Bern}(\exp\{ l_i - C \})$. Although both approaches have the same goal, the Poisson-zeros option is more popular in the literature.

The main conclusion of this section is that the \texttt{pexm} module provides a simpler and more attractive implementation of a Bayesian PE model in \texttt{JAGS}, since it avoids the use of alternative strategies to indirectly specify the likelihood. In the next section, we develop two types of studies using artificial data sets. The first one compares the results from the model in Code chunk CC.2 and those obtained via the \texttt{msm} package. The second study is intended to evaluate the performance of \texttt{pexm} with respect to the Poisson-zeros strategy. In Section \ref{secresult}, the analysis ``\texttt{pexm} versus zeros-trick'' is based on a real data set and it accounts for scripts (zeros-trick case) suggested in \cite{Ibra01}.

%%%%%%%%%%%%%%%%%%%%%%%%%%%%%%%%%%%%%%%%%%%%%%%%%%%%%%%%%%%
\subsection[Simulation study.]{Simulation study.} \label{secsim}
%%%%%%%%%%%%%%%%%%%%%%%%%%%%%%%%%%%%%%%%%%%%%%%%%%%%%%%%%%%

The first goal of the present section is to verify whether the proposed \texttt{JAGS} module is correctly implemented and working as expected. In order to carry out this type of analysis, we consider the configuration described in Section \ref{secR} to generate a sample of $1{,}000$ time points from the PE distribution; the function \texttt{rpexp} (package \texttt{msm}) is applied here. In each interval defined by $\tau$, we have observed the following number of time points: 443 in $I_1$, 247 in $I_2$, 245 in $I_3$ and 65 in $I_4$. 

\begin{figure}[!h]
\centering
 \mbox{\small (a)} \\
 \includegraphics[width=0.80\textwidth]{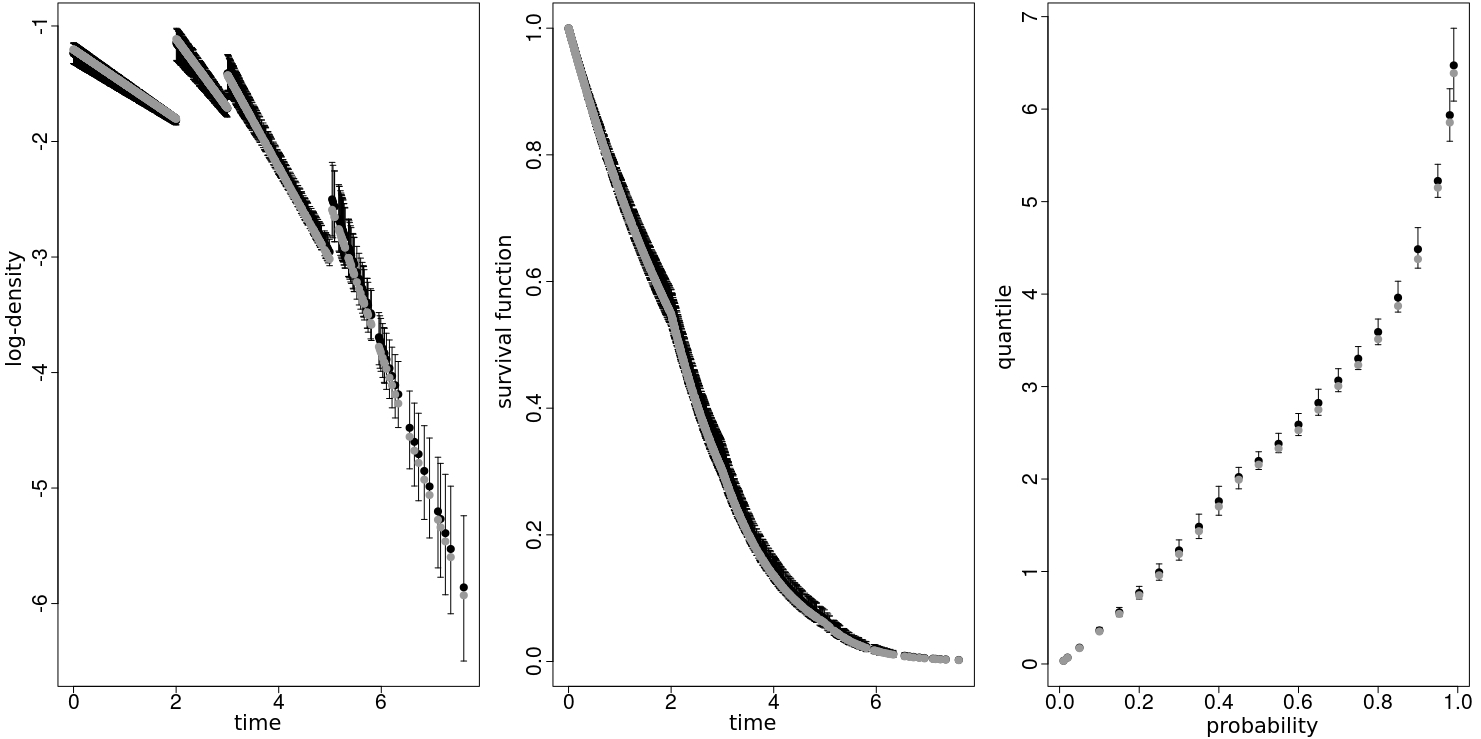} \vspace{3pt} \\
 \mbox{\small (b)} \\ 
 \includegraphics[width=0.80\textwidth]{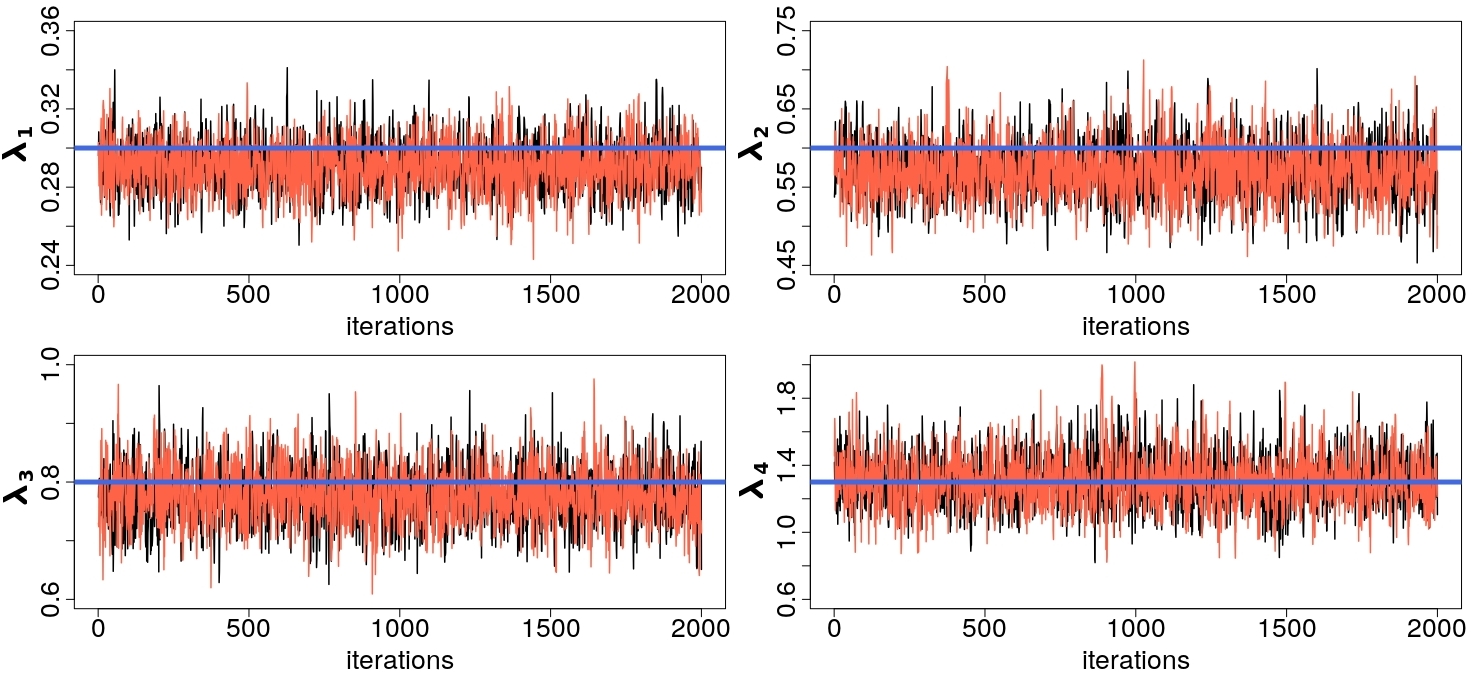} \\
\vspace{-10pt}
\caption{Comparison of posterior estimates (mean and $95\%$ HPD interval) obtained from \texttt{JAGS} (black) and the values calculated via \texttt{msm} package (grey points). Panel (a): log-density and survival function for each observation $t_i$ and quantiles for different probabilities. Panel (b): trace plots representing the two chains (in red and black after the burn-in period) for each $\lambda_j$; horizontal lines (blue) indicate the real values.}
\label{f_pexm_msm}
\end{figure}

The \texttt{R} script shown in Code chunk CC.1 is used to run the MCMC for the simple PE model indicated in Code chunk CC.2. The argument \texttt{data} is set to evaluate \texttt{nq = 23} different quantiles corresponding to probabilities in \texttt{pq} ranging from $0.01$ to $0.99$. A good behavior is confirmed, if all posterior estimates for $\lambda$ are close to their corresponding real values (see Section \ref{secR}) and the chain built for each log-density, survival function and quantile is centered around the true value calculated through one of the existing PE functions of the package \texttt{msm}. Figure~\ref{f_pexm_msm} (a) clearly indicates this mentioned configuration. The three graphs show all true values (grey points) inside the $95\%$ HPD (Highest Probability Density) intervals obtained based on the results from \texttt{JAGS}. In this paper, all HPD intervals are calculated through the \texttt{R} package \texttt{coda}. Other interesting aspects can be observed in Panel (a): ($i$) the shape and decay speed of the survival function; ($ii$) the log density formed by four different parts corresponding to each interval $I_j$; ($iii$) the higher posterior uncertainty expressed in the log-density related to $I_4$. The behavior indicated in ($iii$) is justified by the fact that only 65 time-points belong to the last grid interval, meaning less information to estimate $\lambda_4$.

Figure~\ref{f_pexm_msm} (b) shows the chains generated for each $\lambda_j$ after the burn-in period. As it can be seen, the convergence to the real values (horizontal lines) is visually confirmed and low autocorrelation is observed for all cases (recall that \texttt{lag = 1}). Here, all real values can be found within the corresponding $95\%$ HPD intervals, and the posterior mean and median approximate well their targets. In terms of variability, the standard deviations can be ordered as follows: $0.013 \, (\lambda_1) < 0.037 \, (\lambda_2) < 0.050 \, (\lambda_3) < 0.165 \, (\lambda_4)$. Again, a larger number of observations in $I_j$ leads to a smaller posterior variability for $\lambda_j$.

The Code chunk CC.2 contains the \texttt{pexm} functions \texttt{hpex} and \texttt{hcpex}, which are alternatives to calculate during the MCMC run the hazard function (\ref{eqht}) and the cumulative hazard function (\ref{eqHt}), respectively. Note that this \texttt{JAGS} model script is set to evaluate these functions for $t_{100} = 3.483$. Since $\tau = \{0, 2, 3, 5\}$, we have $t_{100} \in I_3$ and thus $h(t_{100}) = \lambda_3$. The posterior sample related to the object \texttt{ht100} provides the mean $0.778$ (the true $\lambda_3$ is $0.8$). The posterior mean of the cumulative hazard \texttt{Ht100} is $1.529$ (the true value is $1.587$). This average can also be obtained using the relationship $H(t) = -\ln[S(t)]$. A final remark related to these functions is that the present example can be extended to the survival regression setting. One might be interested in evaluating $h(t)$ or $H(t)$ under the effect of some fixed configuration of covariates. This can be easily adapted to \texttt{hpex} and \texttt{hcpex} by including the covariates effects in the argument \texttt{lambda}; Section \ref{secresult} discusses two models having a regression setting.  

\vspace{5pt}
\noindent \textbf{Monte Carlo scheme.} 
\vspace{5pt}

The second investigation developed in this section is based on a Monte Carlo (MC) scheme with 100 replications, i.e., we evaluate 100 data sets generated under the same conditions. The main aim here is to compare the performances of a Bayesian model implemented with the module \texttt{pexm} and another version using the Poisson-zeros strategy (zeros-trick). We emphasize that the difference here is only the \texttt{JAGS} syntax. The model to be fitted is a simplification of the one exhibited in the Code chunk CC.2. We will not save results related to the quantiles, survival, hazard and cumulative hazard functions. In the \texttt{pexm} case, we only specify the likelihood via \texttt{t[i] ~ dpex(lambda[], tau[])} and the prior \texttt{lambda[j] ~ dgamma(0.01, 0.01)}. On the other hand, the Poisson-zeros case is defined as indicated in Code chunk CC.3, where the log-likelihood related to the PE distribution must be written for the object \texttt{l[i]}; we also set $\lambda_j \sim \mbox{Ga}(0.01, 0.01)$. Using this simple model, we evaluate ``\texttt{pexm} versus zeros-trick'' based on the MCMC outputs associated to the rates $\lambda_j$'s.  

In order to generate the data sets, consider the steps: ($i$) set the grid $\tau = \{0, 2, 3, 5\}$ establishing 4 intervals, ($ii$) choose the sample size $n$, ($iii$) define the true $\lambda$ and ($iv$) generate time observations from the PE distribution using the function \texttt{rpexp} from the \texttt{R} package \texttt{msm}. In the present analysis, we explore three scenarios of configurations for the true $\lambda$: increasing $\lambda = \{0.3, 0.6, 0.8, 1.3\}$ (Scenario 1 or S1), constant $\lambda = \{0.7, 0.7, 0.7, 0.7\}$ (Scenario 2 or S2) and decreasing $\lambda = \{1.3, 0.8, 0.6, 0.3\}$ (Scenario 3 or S3). In addition to these scenarios, we evaluate two sample sizes ($n = 100$ and $n = 1{,}000$ time points).

For both implementations, the MCMC is configured with burn-in $= 1{,}000$, lag $= 1$ and $2{,}000$ posterior samples collected after the warm-up period. Two chains are obtained for each parameter. These chains are combined to calculate all descriptive statistics; except the effective sample size (using only the first chain). The starting values in the algorithm are also the same for both strategies; we highlight that the arguments \texttt{.RNG.name} and \texttt{.RNG.seed}, see the \texttt{rjags} documentation, were fixed to impose the same settings of random number generators in \texttt{JAGS}. The algorithms for all combinations of scenarios and sample sizes were executed in the same computer (Intel Core i7 processor and 16.00 GB memory) dedicated exclusively for this task. 

\vspace{-10pt}

\begin{figure}[!h]
\centering
$$
 \begin{array}{ccc} 
  (\mbox{a}) & \hspace{0.5cm} & (\mbox{b}) \\
  \includegraphics[width=0.36\textwidth]{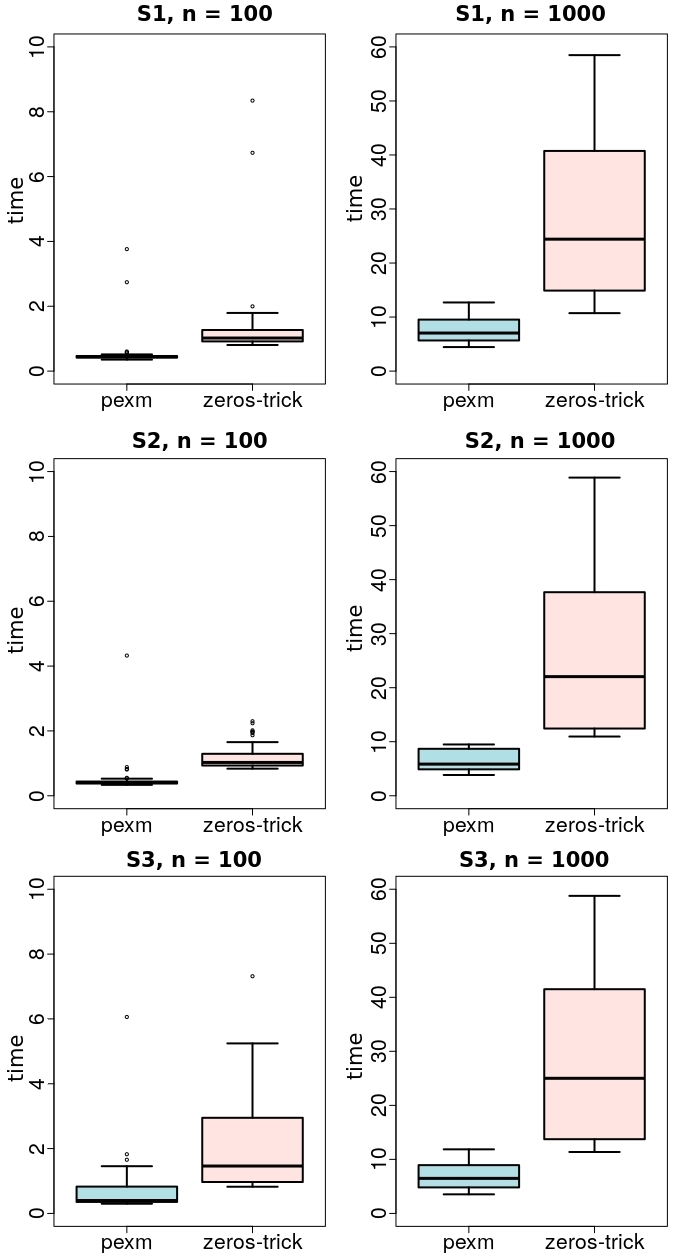} & \hspace{0.5cm} &
  \includegraphics[width=0.44\textwidth]{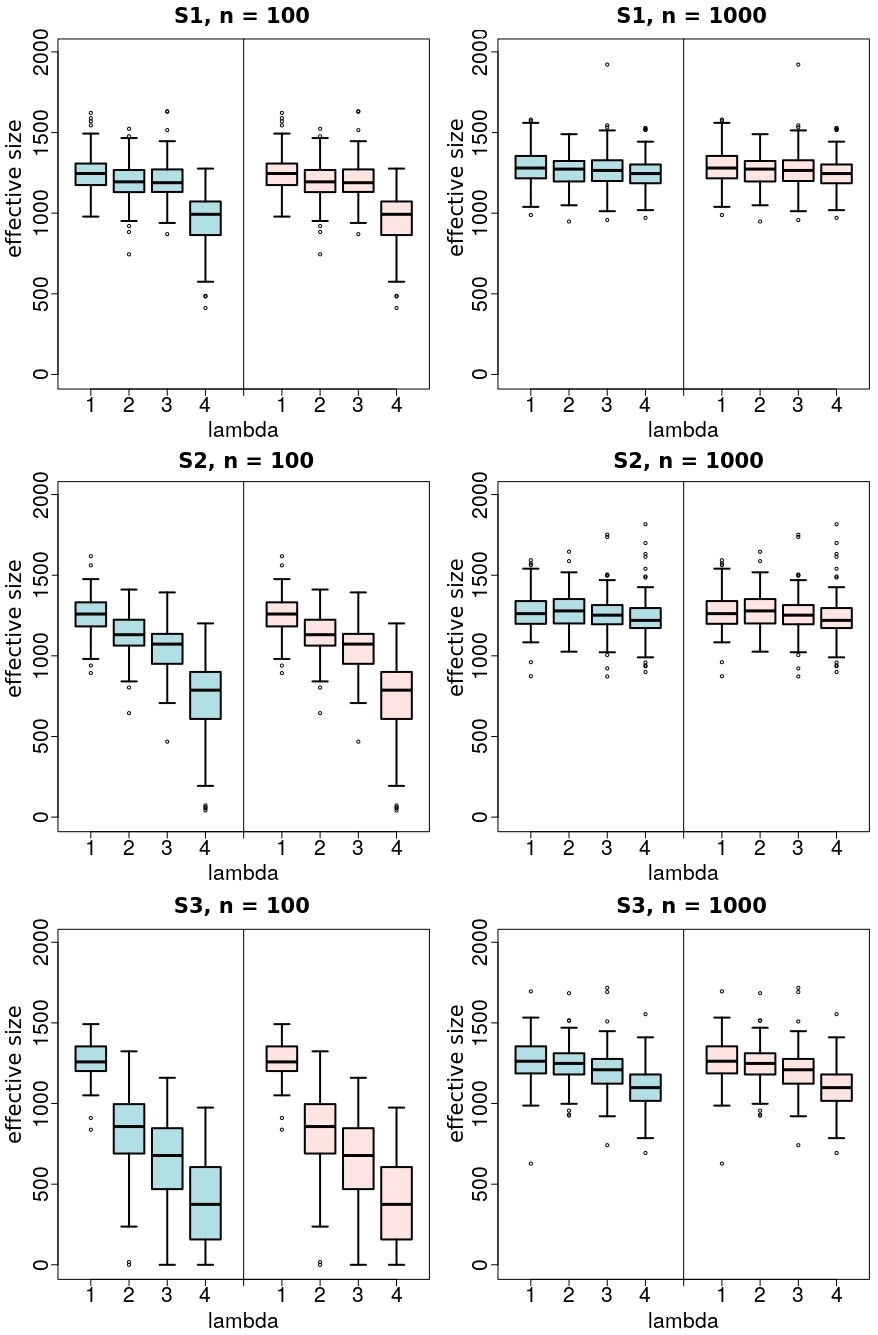} \\
 \end{array}
$$
\vspace{-15pt}
\caption{Boxplots summarizing the results from the MC scheme. Graphs in light blue (left) are related to the implementation via \texttt{pexm}. Graphs in light red (right) correspond to the Poisson-zeros strategy. Panel (a) shows the computational times required by \texttt{JAGS} to run the MCMC (Column 1 indicates $n = 100$, Column 2 indicates $n = 1{,}000$). Panel (b) presents the effective sample sizes of the chains generated for each $\lambda_j$ (Column 1 indicates $n = 100$, Column 2 indicates $n = 1{,}000$). In both panels, the scenarios S1, S2 and S3 are represented in the rows 1, 2 and 3, respectively.}
\label{f_mcsim}
\end{figure}

Figure~\ref{f_mcsim} shows boxplots summarizing some of the most interesting results from the MC scheme. Thys type of graph is a good option in studies involving replications, since it indicates the variability of repeated measurements (Monte Carlo error) through the length of the boxes. Panel (a) presents the computational times (in seconds) needed to run the Gibbs sampling via \texttt{JAGS}. Theses values were determined through the existing function \texttt{proc.time} in \texttt{R}. The graphs in Panel (b) indicate the effective sample size of the chains obtained for the  $\lambda_j$'s. This quantity was calculated using the function \texttt{effectiveSize} available in the \texttt{R} package \texttt{coda}. The light blue color represents the implementation based on the module \texttt{pexm}; light red is related to the Poisson-zeros strategy. Given that the same data set is being fitted by the same model with different implementations, this is done in a controlled setting for reproducibility, we have observed the exact same results from the corresponding MCMC runs. In other words, the outcomes of \texttt{pexm} and zeros-trick are the same in the comparison of bias, coverage percentages and autocorrelation. This aspect is clearly noted in Panel (b). %and other statistics can be inspected through the supporting code in the replication materials of this paper. 

In many studies, the computational time is measured in terms of the processor time used by each implementation to achieve a given effective sample size. This alternative cannot be applied here due to the mentioned compatibility of results. Instead, we focus on the required time to run $3{,}000$ MCMC iterations for each artificial data set. Figure~\ref{f_mcsim} (a) presents the main difference between \texttt{pexm} and the zeros-trick. Naturally, the computational times related to $n = 100$ are smaller than those for fitting $n = 1{,}000$; please, note the distinct scales in the y-axis to allow visualization. For all combinations (scenario, $n$), the boxplots associated with \texttt{pexm} are located in a lower level than those for the zeros-trick. This suggests that the computational speed is an advantage of the module.  

Figure~\ref{f_mcsim} also indicates that the differences between the scenarios S1, S2 and S3 are more evident for the case $n = 100$. Note that the variability exhibited by the graphs are higher in (S3, $n = 100$). Few distinctions are detected between the scenarios when $n = 1{,}000$. Panel (b) suggests that the effective sample size related to $\lambda_1$ tend to be higher than those for the other rates. Recall, from the previous study, that the true $\lambda$ in S1 provides few time points generated in the last interval of the grid ($t_i > 5$). This feature also occurs for the choices of $\lambda$ in S2 and S3. When comparing the number of cases $t_i > 5$, we have the following order S1 > S2 > S3. The small number of observations in the last interval (especially for S3) explains the worst performance of $\lambda_4$ in terms of effective sample size.  

%%%%%%%%%%%%%%%%%%%%%%%%%%%%%%%%%%%%%%%%%%%%%%%%%%%%%%%%%%%
\section[Real application]{Real application.} \label{secresult}
%%%%%%%%%%%%%%%%%%%%%%%%%%%%%%%%%%%%%%%%%%%%%%%%%%%%%%%%%%%

The main goal of this section is to explore the \texttt{pexm} module in a Bayesian analysis via JAGS for a real data set. Two frailty PE models are considered here; they were proposed in \cite{Sahu97} and reported in \cite{Ibra01} for the kidney catheter data in \cite{Mcgil91}. The comparison ``\texttt{pexm} versus zeros-trick'' is also developed in this analysis.   

The kidney catheter is a well known data set often used to illustrate survival models with random effects; it is also available in the \texttt{R} package \texttt{survival} \citep{The15}. The response variable is the time to infection from the insertion of a catheter in a patient using portable dialysis equipment. The time of first and second infections are registered for each patient; there are 38 subjects and thus 76 time measurements for the analysis (18 of them are right-censored). After the occurrence (or censoring) of the first event, enough time is allowed to cure the infection before starting the second insertion. The data set includes an indicator variable identifying the event status ($1$ for infection, $0$ for censoring) and three covariates: gender, age in years and disease type (with four categories). Following \cite{Ibra01}, the covariate ``disease type'' is not included in our study.  

Let $T_{ik}$ be the random variable representing the infection time for the $i$-th patient and the $k$-th insertion; $i = 1,\ldots,38$ and $k = 1, 2$. In addition, consider the fixed covariate vector ${\bm x}_{ik} = (x_{\sex i}, x_{\age ik})^\top$ where $x_{\sex i}$ is the gender ($1 =$ female, $0 =$ male) and $x_{\age ik}$ is the age of the $i$-th patient in the $k$-th insertion. Given ${\bm x}_{ik}$ and the unobserved positive random variable $z_i$ (frailty), the conditional hazard function of $T_{ik}$ in a shared frailty model can be written as:
\begin{equation}
  h(t_{ik}|{\bm x}_{ik}, z_i) \; = \; h_0(t_{ik}) \; \exp\{ {\bm x}_{ik}^\top \beta \} \; z_i, 
  \label{htij}
\end{equation}
with $\beta = (\beta_{\sex},\beta_{\age})^\top$ and $h_0(.)$ being the unknown baseline hazard function for all patients. The non-informative censoring mechanism is assumed in this application. Note that the random effect $z_i$ establishes a dependency between the first and second infection times measured for the same patient. This modeling structure is common in the literature and it can be seen as an extension to the Cox proportional hazards model \citep{Cox72}. Frailty models were first motivated by \cite{Vau79} and then developed by \cite{Cla85}, \cite{Oak86} and \cite{Oak89}.

The PE distribution, discussed in Section \ref{secped}, can be used to represent the baseline hazard in (\ref{htij}); that is $h_0(t_{ik}) = \lambda_j$ for $t_{ik} \in I_j$, $j = 1, \ldots, m$. For simplicity, one can assume a prior configuration treating the $\lambda_j$'s as independent; however, \cite{Sahu97} and \cite{Asl98} argue that in a frailty model the ratio of marginal hazards is similar to the ratio of baseline hazards for near time points (given the same covariates). In this case, it would be more appropriate to consider a prior that correlates the $\lambda_j$'s in adjacent intervals. Following \cite{Ibra01}, we shall explore two approaches:
\begin{itemize}
  \item Model I: \; $(\lambda_j | \lambda_{j-1}) \sim \mbox{Ga}(\alpha_j, \, \alpha_j/\lambda_{j-1})$ with $\lambda_0 = 1$; \vspace{-3pt}
  \item Model II: \; $\xi_j = \ln(\lambda_j)$ and $(\xi_j | \xi_{j-1}) \sim N(\xi_{j-1}, \nu)$ with $\xi_0 = 0$.
\end{itemize}

In the structure of both models, we also have: $z_i \sim \mbox{Ga}(\eta,\eta)$, $\eta \sim \mbox{Ga}(\phi_1,\phi_2)$ and $\beta \sim N_2[{\bm 0}, \mbox{diag}(\sigma^2_1, \sigma_2^2)]$. Denote $\kappa = 1/\eta = \mbox{Var}(z_i)$; large values of $\kappa$ imply in great heterogeneity between patients. We consider $\phi_1 = \phi_2 = 10^{-3}$ and $\sigma^2_1 = \sigma^2_2 = 10^3$, which determines vague prior distributions (variance $1{,}000$) for $\eta$, $\beta_{\sex}$ and $\beta_{\age}$. In addition, let $\alpha_j = 10^{-2} \; \forall \, j$ in Model I and $\nu = 10^4$ in Model II. The reader should refer to \cite{Ibra01} for other comments related to these hyperparameters choices. 

The authors in \cite{Ibra01} also discuss a sensitivity analysis comparing $m = 5$, $10$ and $20$ intervals partitioning the time axis in the PE distribution. They indicate that $m = 5$ provides the worst model fit, and the results for $m = 20$ are not substantially better than $m = 10$; therefore, for parsimony, $m = 10$  is the choice selected for this analysis.  

Due to the presence of right-censored observations, we use the standard strategy in \texttt{JAGS} based on the existing \texttt{dinterval} distribution to input the missing infection times; other details can be found in Section A.1, Appendix A. The \texttt{JAGS} scripts based on \texttt{pexm} for the models I and II are presented in the next three code chunks identified as CC.4 (a), (b) and (c). The code in (a) shows the main script containing: the time grid (line 1), the likelihood function (lines 3--12) and the prior distributions (line 13--18) of the frailties and the regression coefficients. Recall that the precision is the second parameter of the normal distribution, according to the \texttt{JAGS} parameterization. Note that the code in line 1 provides $\tau = \{0, 56.2, 112.4, 168.6, 224.8, 281.0, 337.2, 393.4, 449.6, 505.8 \}$, which is an equally spaced time grid built with respect to the largest infection time (562) in the data set. The Code chunks CC.4 (b) and CC.4 (c) show the commands related to the prior distributions correlating the $\lambda_j$'s in models I and II, respectively. Simply insert at the end of CC.4 (a) the script exhibited in (b) or (c) for the full model specification. The \texttt{JAGS} scripts for the PE models based on the zeros-trick are presented in Appendix B.    

\vspace{5pt}
\noindent \textbf{Code chunk CC.4 (a)} 
\vspace{-5pt}
\begin{lstlisting}[numbers=left]
data { for(j in 1:m) { a[j] <- 562 * (j - 1) / m } } 
model{  
  for (i in 1:n){  
    for(k in 1:2){
      censored[i, k] ~ dinterval( t[i, k], t_cen[i, k]) 
      t[i, k] ~ dpex(haz[i, k, ], a[])
      for(j in 1:m){ 
        haz[i, k, j] <- lambda[j] * exp( beta_sex * sex[i] + beta_age * age[i, k] ) * z[i] 
        }
     }
  }
  for(i in 1:n){ z[i] ~ dgamma(eta, eta) }
  kappa <- 1 / eta
  eta ~ dgamma(0.001, 0.001)
  beta_sex ~ dnorm(0, 0.001)
  beta_age ~ dnorm(0, 0.001)
  ... # Include here the Code chunk CC.4 (b) or (c).
}
\end{lstlisting}

\vspace{5pt}
\noindent \textbf{Code chunk CC.4 (b)} 
\vspace{-5pt}
\begin{lstlisting}[numbers=left]
  rate[1] <- 0.01 / 1
  lambda[1] ~ dgamma(0.01, rate[1])
  for(j in 2:m){
    rate[j] <- 0.01 / lambda[j - 1]
    lambda[j] ~ dgamma(0.01, rate[j])
  } 
\end{lstlisting}

\vspace{5pt}
\noindent \textbf{Code chunk CC.4 (c)} 
\vspace{-5pt}
\begin{lstlisting}[numbers=left]
  xi[1] ~ dnorm(0, 0.0001)
  lambda[1] <- exp(xi[1])
  for(j in 2:m){
    xi[j] ~ dnorm(xi[j - 1], 0.0001)
    lambda[j] <- exp(xi[j])
  } 
\end{lstlisting}

We can handle right-censored observations in \texttt{JAGS} by defining three variables to be inserted as data: ($i$) a binary indicator \texttt{isCensored} with $1$ for censored values, ($ii$) a time response vector (say \verb't_res') with ``NA'' replacing the censored cases and ($iii$) a censored time vector (say \verb't_cen') having $0$ for the non-censored subjects. The core of the \texttt{JAGS} model for censored data is represented by the two commands: \verb'isCensored[i] ~ dinterval(t_res[i], t_cen[i])' and \verb't_res[i] ~ dpex(lambda[], tau[])'. Here, \texttt{JAGS} automatically imputes a random value for a missing observation \verb't_res[i] = NA'. The corresponding response \texttt{isCensored[i] = 1} determines that the observation to be imputed must be greater than \verb't_cen[i]'; this is implied by the existing \texttt{JAGS} distribution \texttt{dinterval} \citep[see][]{Plu03}. In this example, the new observation is then generated from the $\mbox{PE}(\lambda, \tau)$ with lower bound \verb't_cen[i]' and unchanged upper bound $+\infty$. 

The MCMC configuration for the present study is as follows: burn-in $= 10{,}000$ iterations, lag $= 1$, posterior sample composed by $10{,}000$ observations and two chains being generated for each parameter. In terms of initial values, we set for both models the arguments: \texttt{.RNG.name = "base::Super-Duper"} and \texttt{.RNG.seed = 5} for the first chain and \texttt{.RNG.name = "base::Wichmann-Hill"} and \texttt{.RNG.seed = 2} for the second chain. See the \texttt{rjags} documentation \citep{Plu16} for further details about starting values in this tool.

\begin{table}[!h]
\centering \scriptsize
\begin{tabular}{c|cc|rrrc}
\hline
\multicolumn{1}{r}{} & Model & Script & Mean & Median & S.D. & HPD \\ 
\hline
& I  & Module  & -1.4727 & -1.4669 & 0.4888 & (-2.3964, -0.4987) \\ 
$\beta_{\sex}$ &    & Poisson & -1.4764 & -1.4639 & 0.4648 & (-2.4195, -0.6104) \\ \cline{2-7}
& II & Module  & -1.4593 & -1.4399 & 0.4675 & (-2.3901, -0.5477) \\ 
&    & Poisson & -1.4650 & -1.4529 & 0.4802 & (-2.3769, -0.5111) \\ 
\hline 
& I  & Module  &  0.0076 &  0.0071 & 0.0123 & (-0.0165, 0.0319) \\ 
$\beta_{\age}$ &    & Poisson &  0.0055 &  0.0051 & 0.0115 & (-0.0168, 0.0290) \\ \cline{2-7}
& II & Module  &  0.0072 &  0.0070 & 0.0116 & (-0.0161, 0.0295) \\ 
&    & Poisson &  0.0063 &  0.0064 & 0.0123 & (-0.0174, 0.0304) \\ 
\hline 
& I  & Module  &  0.5043 &  0.4642 & 0.2774 &  (0.0388, 1.0271) \\ 
$\kappa$ &    & Poisson &  0.4924 &  0.4507 & 0.2664 &  (0.0242, 0.9939) \\ \cline{2-7}
& II & Module  &  0.4838 &  0.4430 & 0.2739 &  (0.0350, 0.9937) \\ 
&    & Poisson &  0.4968 &  0.4502 & 0.2823 &  (0.0286, 1.0261) \\    
\hline   
\end{tabular}
\caption[]{Posterior estimates and 95\% HPD intervals for regression coefficients ($\beta$) and variance ($\kappa$) of the frailties. Comparisons: ``Model I vs. II'' and ``\texttt{pexm} vs. zeros-trick''. \label{t_descrip}}
\end{table}

Table~\ref{t_descrip} presents the posterior mean, median, standard deviation and 95\% HPD credible intervals for $\beta_{\sex}$, $\beta_{\age}$ and $\kappa$. Here, we can compare the results from models I and II implemented with \texttt{pexm} and the Poisson-zeros strategy. In contrast with the simulation study, we do not observe here the exact same result when comparing the same model and parameter for both implementation types. However, the estimates obtained in this comparison are very similar; in some cases the distinction occurs in the third decimal place. This small difference can be justified by numerical approximation errors affecting the computations related to the more complex models under investigation in this section. Another interesting aspect to be highlighted is the similarity between the estimates from models I and II. As a result, one cannot choose one of the models based on a more reasonable interpretation of a parameter. Note that all credible intervals for $\beta_{\age}$ include zero, indicating that the age has no significant effect over the infection time. On the other hand, all intervals for $\beta_{\sex}$ are below zero, meaning that a female patient has lower infection risk than a male patient for a given time point. 

According to \cite{Ibra01}, the posterior means of $\kappa$ in Table~\ref{t_descrip} should be considered high, providing evidence of a strong positive association between two infection times for the same patient ($T_{i1}$ and $T_{i2}$). A large $\kappa$ suggests that a major part of the variability in the data is explained by the clusters (the patients) rather than the insertions for the same individual.  

\begin{figure}[!h]
\centering
  \includegraphics[width=0.90\textwidth]{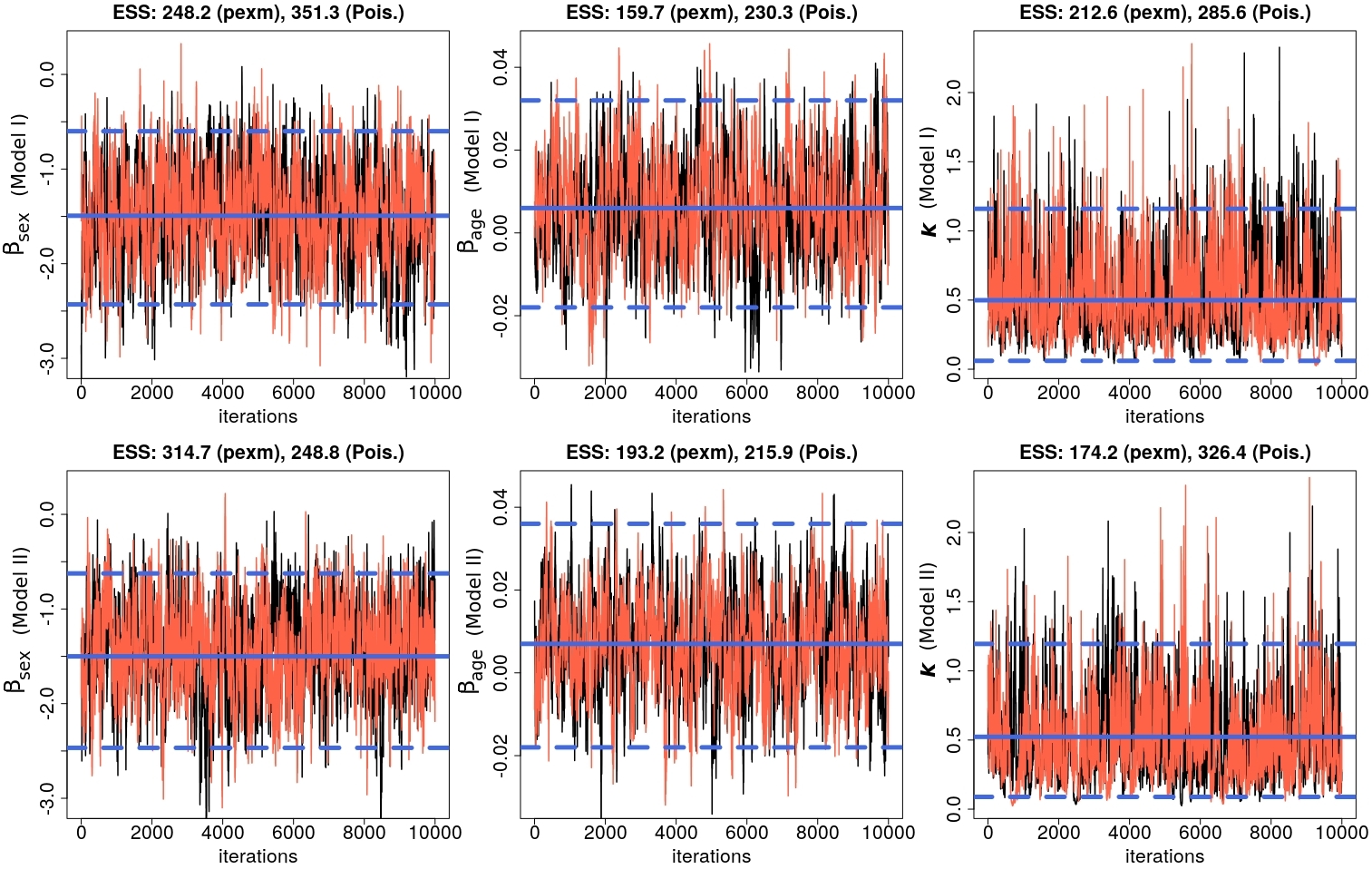} 
\vspace{-10pt}
\caption{Trace plots representing the first chain (after burn-in period) for $\beta$ and $\kappa$. The posterior samples related to \texttt{pexm} are in red and those related to the Poisson-zeros are in black. The horizontal lines indicate the estimates reported in \cite{Ibra01}: mean (continuous) and 95\% credible intervals (dashed). Effective sample sizes (ESS) are presented at the top of each graph.}
\label{f_chains}
\end{figure}

The similarity of the estimates in Table~\ref{t_descrip}, for the comparison ``\texttt{pexm} versus zeros-trick'', suggests that the chains are converging to the same region in the parameter space and exhibiting equivalent variability. This behavior is expected since we fit the same model. The trace plots in Figure~\ref{f_chains} confirm this interpretation and also indicate that the level of autocorrelation obtained via \texttt{pexm} tends to reflect the one produced via the Poisson-zeros strategy. The effective sample sizes of the chains, closely related to the autocorrelations, are shown at the top of the graphs. The values are not high, suggesting some strong autocorrelation that can be reduced by setting \texttt{lag} > 1 in the MCMC configuration. Although small ESS are reported, great discrepancies are not detected when comparing the implementations. The three horizontal lines, included in each graph, represent the mean and the 95\% credible interval reported in \cite{Ibra01} for the corresponding parameter. Note that all chains are converging to the same region indicated in the literature. 

\begin{figure}[!h]
\centering
  \includegraphics[width=0.75\textwidth]{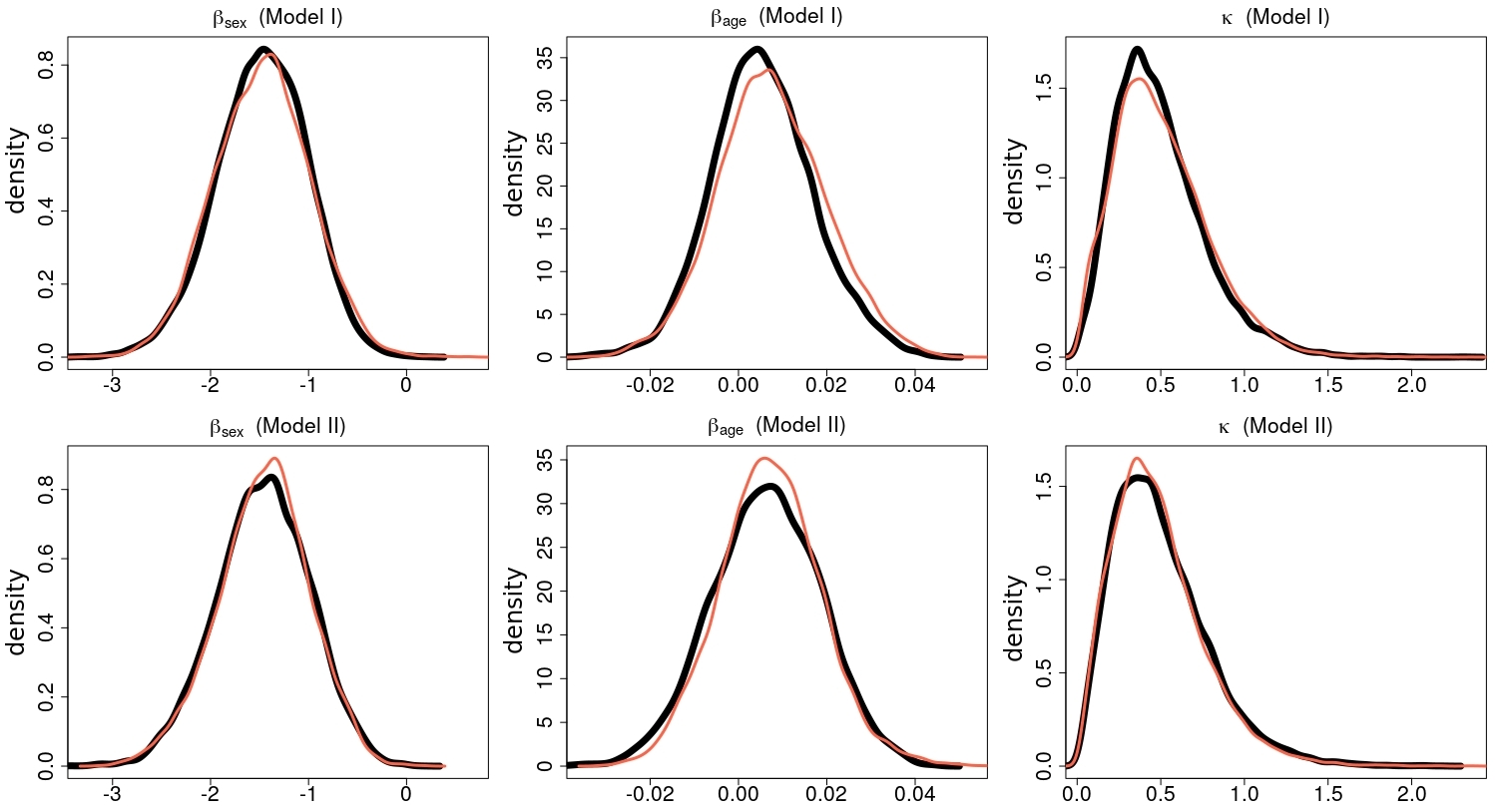} 
\vspace{-10pt}
\caption{Kernel density estimation (assuming the default ``Gaussian'' smoothing kernel in \texttt{R}) for the posterior samples of $\beta$ and $\kappa$. Results using the module are in red and the Poisson-zeros strategy in black.}
\label{f_dens}
\end{figure}

\begin{figure}[!h]
\centering
  \includegraphics[width=0.75\textwidth]{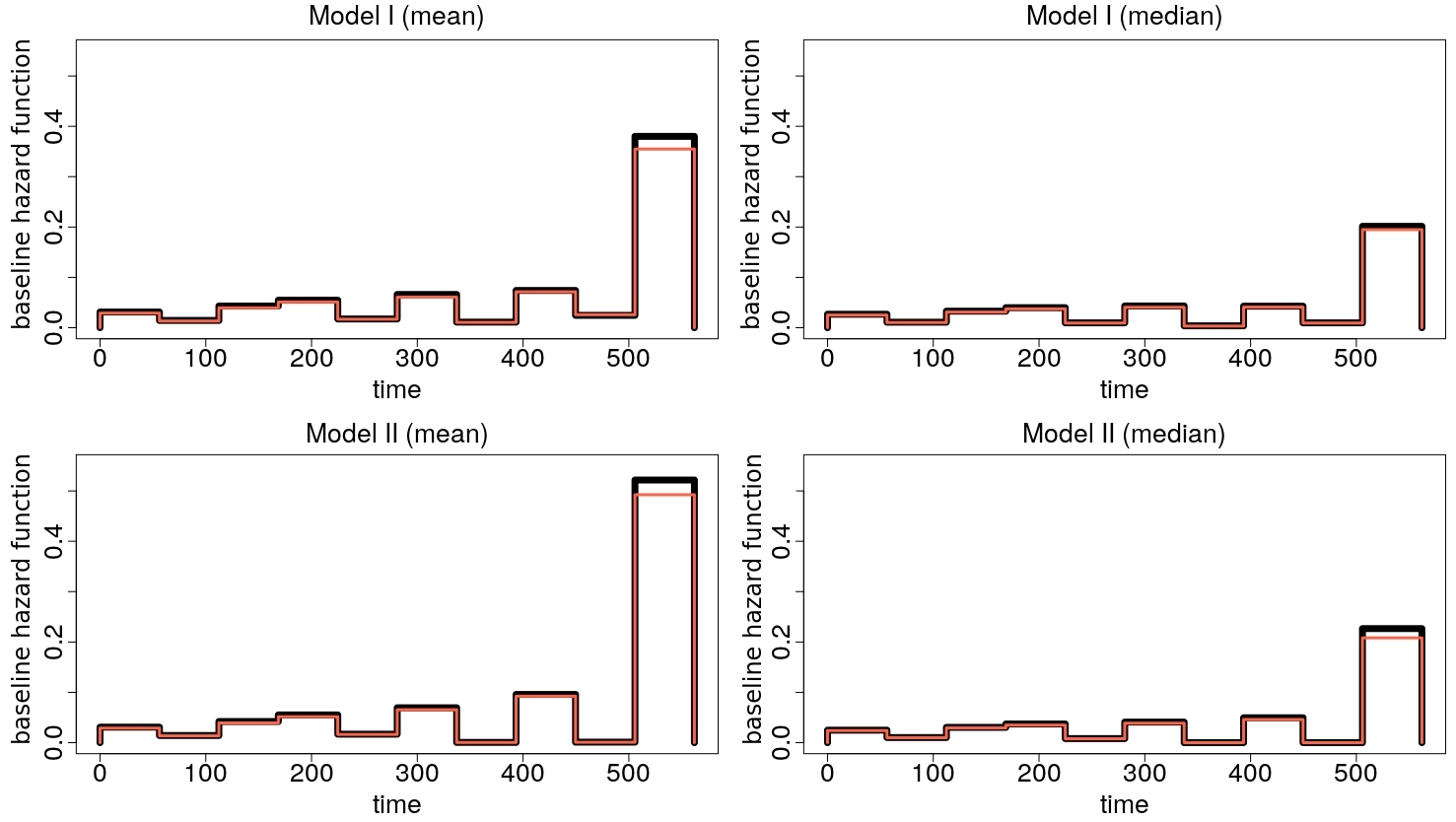} 
\vspace{-10pt}
\caption{Comparison of posterior estimates for each $\lambda_j$ indexing the PE distribution. Consider: \texttt{pexm} (in red), zeros-trick (in black), mean (first column), median (second column), Model I (first row) and Model II (second row).}
\label{f_basehaz}
\end{figure}

Figure~\ref{f_dens} shows the estimated kernel densities obtained in \texttt{R} for the posterior samples generated in this real application. Note that the curves in red (PE module) and black (Poisson-zeros) can be considered similar in all cases, which reinforces the idea of two algorithms working as expected to fit the same model. Very few discrepancies are detected in these graphs, especially around their modes; however, this can be significantly reduced when assuming a larger posterior sample size (say $100{,}000$ iterations after the burn-in). In Figure~\ref{f_basehaz}, we compare the estimated baseline hazard functions $h_0(t_{ik})$ obtained via \texttt{pexm} (red) and the zeros-trick (black). As it can be seen, the results from both cases tend to agree. A slightly difference is detected in the time interval related to $\lambda_{10}$. Such small distinction almost disappears when evaluating the posterior medians. 

\begin{table}[!h]
\centering \scriptsize
\begin{tabular}{c|c|rrrrrrrrrr}
  \hline
Model & Script & $\lambda_1$ & $\lambda_2$ & $\lambda_3$ & $\lambda_4$ & $\lambda_5$ & $\lambda_6$ & $\lambda_7$ & $\lambda_8$ & $\lambda_9$ & $\lambda_{10}$\\
 \hline
 I  & Module  & 0.0183 & 0.0109 & 0.0312 & 0.0481 & 0.0216 & 0.0623 & 0.0223 & 0.0916 & 0.0479 & 0.4570 \\
    & Poisson & 0.0195 & 0.0122 & 0.0359 & 0.0503 & 0.0234 & 0.0740 & 0.0200 & 0.0910 & 0.0455 & 0.5568 \\
 \hline
 II & Module  & 0.0194 & 0.0134 & 0.0351 & 0.0579 & 0.0296 & 0.1005 & 0.0043 & 0.1764 & 0.0122 & 1.1870 \\
    & Poisson & 0.0222 & 0.0140 & 0.0401 & 0.0629 & 0.0310 & 0.0945 & 0.0058 & 0.1690 & 0.0208 & 1.2393 \\
 \hline
\end{tabular}
\caption[]{Posterior standard deviations for each $\lambda_j$. Comparison: ``Model I vs. II'' and ``\texttt{pexm} vs. zeros-trick''. \label{t_sd_lam}}
\end{table}

Table~\ref{t_sd_lam} presents the posterior standard deviations of each $\lambda_j$. Note that the variability tend to be higher when the index $j$ is large. This aspect is related to the choice of the time grid $\tau$. Assuming $m = 10$ and the equally spaced configuration, the number of observed time points (18 censored cases not included) in each interval $I_j$ are: 30, 5, 9, 5, 1, 3, 0, 2, 0 and 3 for $j = 1,\ldots,10$, respectively. Therefore, the last intervals (and their neighbors) have fewer observations, which determines higher posterior uncertainty for the corresponding rates. 

The module \texttt{pexm} allows a simpler implementation of a Bayesian PE model, but this is not the only advantage observed in this paper. The MC study in Section \ref{secsim} clearly indicates that the computational speed is another positive aspect of the proposed tool. The gain in terms of processor time can also be noted when running the MCMC related to the models I and II in the present real illustration. Using the same computer mentioned in the MC study (Section \ref{secsim}), the Poisson-zeros strategy is again considerably slower than the \texttt{pexm} implementation. The function \texttt{proc.time} in \texttt{R} provides the following times (in seconds) to execute the MCMC via \texttt{JAGS}: 24.493 (Model I, PE module), 23.954 (Model II, PE module), 70.857 (Model I, Poisson-zeros) and 68.168 (Model II, Poisson-zeros). 

%%%%%%%%%%%%%%%%%%%%%%%%%%%%%%%%%%%%%%%%%%%%%%%%%%%%%%%%%%%
\section[Conclusions]{Conclusions.} \label{secfinal}
%%%%%%%%%%%%%%%%%%%%%%%%%%%%%%%%%%%%%%%%%%%%%%%%%%%%%%%%%%%

The PE distribution has been widely used in semiparametric settings to model data from applications in survival analysis and reliability. Implementing a PE model is a difficult task for anyone aiming a Bayesian analysis without the necessary programming skills to write all MCMC steps. In order to circumvent this issue, the programs based on the \texttt{BUGS} language are seen as attractive alternatives by many users. However, none of these programs has a module designed to deal with the PE distribution. In this case, a well known implementation strategy is to consider the Poisson or the Bernoulli distributions to indirectly specify the likelihood, which again poses some programming challenges.

\texttt{JAGS} deserves a special attention in the group of programs similar to \texttt{BUGS}, since it is free and extensible. The main goal of this paper was to present a new \texttt{JAGS} module (\texttt{pexm}) to fit Bayesian models accounting for the PE distribution. We have provided a clear picture of how the new module can be used from within the \texttt{R} environment (integrated with \texttt{JAGS}), which is important to guide the interested users. The discussion emphasizes that the module simplifies the \texttt{JAGS} model script. 

A study based on artificial data was developed to compare results using \texttt{pexm} and those from an \texttt{R} package (\texttt{msm}) containing utilities for the PE distribution. The investigation concludes that the module works well. An MC simulation scheme with 100 replications, using a simple Bayesian model, was also conducted to evaluate the behavior of \texttt{pexm} with respect to the usual zeros-trick implementation. The main conclusion here is that the module provides the exact same results with the advantage of having higher computational speed. 

In a real data application, we have considered a data set with two covariates and the  response ``time to infection'' for patients using a portable dialysis equipment. Two PE survival frailty models in the literature were explored in this part. They assume different structures of association between adjacent intervals in the time grid. The presence of right censored data allowed a more in depth evaluation of the PE module; censoring was not considered in the simulation study. Discussions were again devoted to the comparison ``\texttt{pexm} versus zeros-trick''. As expected for two algorithms fitting the same model, their results were quite similar. The small estimation distinctions are possibly due to numerical approximation errors related to complexity of the tested models. In addition, the posterior estimates of the parameters resembles the ones reported in the literature using the same data set.

One last contribution of this work is the compilation of important information for those readers (with some \texttt{C++} knowledge) interested in extending \texttt{JAGS}. The obstacles, solutions and other aspects related to the module implementation are fully described in the Appendix. 

\vspace{10pt}
%%%%%%%%%%%%%%%%%%%%%%%%%%%%%%%%%%%%%%%%%%%%%%%%%%%%%%%%%%%
{\flushleft \large \textbf{Acknowledgements} }
%%%%%%%%%%%%%%%%%%%%%%%%%%%%%%%%%%%%%%%%%%%%%%%%%%%%%%%%%%%
\vspace{5pt}

%The authors would like to thank two anonymous referees for their constructive comments leading to an improved version of this paper and to the \texttt{R} package developed to install the new module. 
The first author also thanks Funda\c{c}\~{a}o de Amparo \`{a} Pesquisa do Estado de Minas Gerais (FAPEMIG) for supporting this research. 

\vspace{10pt}
%%%%%%%%%%%%%%%%%%%%%%%%%%%%%%%%%%%%%%%%%%%%%%%%%%%%%%%%%%%
\renewcommand{\thefigure}{A.\arabic{figure}} \setcounter{figure}{0}
\renewcommand{\theequation}{A.\arabic{equation}} \setcounter{equation}{0}
\renewcommand{\thetable}{A.\arabic{table}} \setcounter{table}{0}
\renewcommand{\thesubsection}{A.\arabic{subsection}}\setcounter{subsection}{0}
{\flushleft \large \textbf{Appendix A: Technical aspects to build the proposed \texttt{JAGS} module.} }
%%%%%%%%%%%%%%%%%%%%%%%%%%%%%%%%%%%%%%%%%%%%%%%%%%%%%%%%%%%
\vspace{10pt}

First, we highlight the fact that this description is based on the work of \cite{Wab14}. This reference is a nice short tutorial indicating the key steps to create a \texttt{JAGS} module. We call the attention for some important points required for the \texttt{pexm} implementation. The reader should refer to the module source code to follow the details reported here. The source code is available in two different repositories:  \vspace{-5pt}
\begin{itemize}
 \item \url{https://github.com/vdinizm/pexm.git},  \vspace{-5pt}
 \item \url{https://sourceforge.net/projects/jags-pexm/files}. \vspace{-5pt}
\end{itemize}

Consider some generic notations related to important directories in the computer: \verb'JAGS_PATH' is the path to the folder where \verb'JAGS' is installed and \verb'PEXM_PATH' is the path to the folder containing the source code of \texttt{pexm}. The \texttt{JAGS} modules are library files located in the \verb'JAGS_PATH/module' directory of the program. Two \texttt{C++} classes must be defined in the module code: one for the module itself and one for a distribution and their related functions (providing the pdf, cdf, quantiles, hazard and cumulative hazard). In our case, the module class file is called \texttt{pexm.cc} and it is located in the subfolder \verb'PEXM_PATH/src' of the project. The first lines of this file references with \texttt{\#include} the header files containing other functions to be used by the module. The first one is the existing \texttt{JAGS} header file \texttt{Module.h}. The second one is the PE distribution class file \texttt{DPex.h}. The remaining cases indicate function class files required to compute the density (\texttt{DPexFun.h}), cdf (\texttt{PPexFun.h}), quantile (\texttt{QPexFun.h}), hazard function (\texttt{HPexFun.h}) and cumulative hazard function (\texttt{HCPexFun.h}) given a time point $t$ or a probability $p$ of interest. These distribution and function class files are described ahead.

Note that we have chosen different names for the distribution class file (\texttt{DPex.h}) and the function class file to compute the pdf (\texttt{DPexFun.h}). This distinction is necessary in the module implementation, otherwise the system will indicate errors when creating the installation files. This is an important aspect to be clarified, because these class files are related to two \texttt{JAGS} model script commands having the same name but applied in different situations. As an illustration, consider: \vspace{-5pt}
\begin{itemize}
 \item \texttt{T} $\sim$ \texttt{dpex(lambda, tau)} is related to \texttt{DPex.h} and indicates that $T \sim \mbox{PE}(\lambda, \tau)$; \vspace{-5pt}
 \item \texttt{Z <- dpex(t, lambda, tau)} is related to \texttt{DPexFun.h} and it allocates to the object \texttt{Z} the pdf of the $\mbox{PE}(\lambda, \tau)$ evaluated at the time point $t$. \vspace{-5pt}
\end{itemize}
This equal name configuration is not mandatory, but it is in accordance with the naming standards adopted for the existing distributions available in \texttt{JAGS}.

\subsection{The distribution class files.} \label{secA1}

The file \verb'PEXM_PATH/src/distributions/DPex.h' contains the class prototype of the PE distribution. Near the top of the code, the \texttt{JAGS} parent class \texttt{VectorDist} is specified as the base class. This choice is appropriate for those distributions parameterized by vectors, which is the case of the $\mbox{PE}(\lambda,\tau)$. The examples in \cite{Wab14} are associated with distributions taking scalar quantities as input arguments and thus related to the alternative parent class \texttt{ScalarDist}. The process of building a \texttt{JAGS} module is easier for scalar valued distributions as the \texttt{RScalarDist} class can be inherited from, which partly automates the availability of \texttt{d}/\texttt{p}/\texttt{q} functions.

The following specific functions were implemented in the \texttt{pexm} \texttt{C++} classes: \vspace{-5pt}
\begin{itemize}
  \item \texttt{logDensity}: calculates the log-density; \vspace{-5pt}
  \item \texttt{randomSample}: generates random samples; \vspace{-5pt}
  \item \texttt{typicalValue}: returns a typical value in the support of the distribution; \vspace{-5pt}
  \item \texttt{checkParameterValue}: indicates if $\{\lambda, \tau \}$ are in the allowed parameter space; \vspace{-5pt}
  \item \texttt{canBound}: indicates if the distribution can be bounded; \vspace{-5pt}
  \item \texttt{isDiscreteValued}: indicates if the PE distribution is discrete (false); \vspace{-5pt}
  \item \texttt{isSupportFixed}: indicates if the upper/lower limits of the PE distribution support are fixed (true, its support does not depend on $\lambda$ or $\tau$); \vspace{-5pt}
  \item \texttt{checkParameterLength}: indicates if $\lambda$ and $\tau$ have the same length; \vspace{-5pt}
  \item \texttt{support}: returns the unbounded support ($0, +\infty$) of the PE distribution.
\end{itemize}
Some of these functions are not required when using the simpler \texttt{ScalarDist} parent class; see \cite{Wab14} for this and other details related to the code.

The file \verb'PEXM_PATH/src/distributions/DPex.cc' contains the actual implementation of the PE distribution. It includes the codes for the nine mentioned specific functions. In the beginning, note that we quote the name \texttt{dpex} and the number of parameters (i.e. 2) to be used in any \texttt{JAGS} script. The implementation of \texttt{logDensity}, \texttt{randomSample} and \texttt{typicalValue} takes into account the elements described in Section \ref{secped}. In particular, \texttt{typicalValue} returns the median of the PE distribution, which is simpler to implement and uses the same structure presented ahead for the quantile function.

The function \texttt{randomSample} has two conditions related to the lower and upper bounds of the PE distribution. The default support is ($0, +\infty$), however, this domain can be bounded to allow applications involving censored data (a very common situation in survival analysis and reliability). Given the existence of new boundaries, the commands implemented within \texttt{if(lower)\{...\}} and \texttt{if(upper)\{...\}} will replace the default interval ($0, 1$) by a shorter one, where a random number is uniformly generated and then transformed into a time point via cdf inversion. The specific function \texttt{canBound} is necessary to enable this restriction.

The function \texttt{checkParameterValue} allows \texttt{JAGS} to alert the user about the wrong specification of $\lambda$ or $\tau$. The following points are verified: ($i$) $\lambda_j \geq 0$ for all $j$, ($ii$) $a_1 = 0$ in $\tau$, ($iii$) $a_1 < a_2 < \ldots < a_m$ in $\tau$. When running \texttt{JAGS}, a warning message will be displayed if at least one of these items is false.   

\subsection{The function class files.} \label{secA2}

The additional functions related to the $\mbox{PE}(\lambda,\tau)$ are located in \verb'PEXM_PATH/src/functions' within the project. The PE class prototype files associated with the pdf, cdf, quantile, hazard and cumulative hazard functions are denoted by \texttt{DPexFun.h}, \texttt{PPexFun.h}, \texttt{QPexFun.h}, \texttt{HPexFun.h} and \texttt{HCPexFun.h}, respectively. Their structures are quite similar, changing only the names from one case to the other. Near the top of each code, we set the \texttt{JAGS} function base \texttt{ScalarVectorFunction.h} allowing the simultaneous specification of one scalar (time $t$ or probability $p$) and two vectors ($\lambda$ and $\tau$) as input arguments.

The directory \verb'PEXM_PATH/src/functions' also contains the ``\texttt{.cc}'' files (they are \texttt{DPexFun.cc}, \texttt{PPexFun.cc}, \texttt{QPexFun.cc}, \texttt{HPexFun.cc} and \texttt{HCPexFun.cc}), where the actual implementations of the mentioned functions are found. The \texttt{JAGS} script names to call these functions (\texttt{dpex}, \texttt{ppex}, \texttt{qpex}, \texttt{hpex} and \texttt{hcpex}) and their number of input arguments (three) are defined at the beginning of each code. As an example, in a \texttt{JAGS} model script the user can allocate the corresponding outcome to the generic objects \texttt{Z1}, \texttt{Z2}, \texttt{Z3}, \texttt{Z4} and \texttt{Z5} as follows:
\begin{lstlisting}[numbers=left]
Z1 <- dpex(t, lambda, tau) 
Z2 <- ppex(t, lambda, tau) 
Z3 <- qpex(p, lambda, tau)
Z4 <- hpex(t, lambda, tau)
Z5 <- hcpex(t, lambda, tau) 
\end{lstlisting}
Note that the first argument is always the scalar ($t$ or $p$), the second one is $\lambda$ and the third one is the grid $\tau$.

\subsection{Installing the PE module.} \label{secA3}

In order to configure the new \texttt{JAGS} module in the local computer, the user must simply install the \texttt{R} package \texttt{pexm} designed to handle this task in Unix-like (Linux or Mac) or Windows systems. See the repositories mentioned in Section \ref{secA1} to access the source code or the package tarball (\texttt{tar.gz} file) related to \texttt{pexm}. In the development of the proposed package, we considered as examples the source codes from \texttt{runjags} \citep{Den16} and \texttt{rjags} \citep{Plu16}, which contain key details about including modules to be used through the interface between \texttt{R} and \texttt{JAGS}. The \texttt{pexm} can be seen as an additional option in this group of examples to guide future developers. Code chunk CC.5 shows three possibilities to request the \texttt{pexm} installation in the \texttt{R} console. We highlight two important notes here: ($i$) the package \texttt{devtools} \citep{devtools} is needed to deal with the option involving the \texttt{GitHub} repository and ($ii$) the appropriate \texttt{tar.gz} file, named with a system indication (\texttt{unix} or \texttt{windows}), should be downloaded into the computer when using the \texttt{Sorceforge} repository. The reader must have in mind that a recent version of the program \texttt{JAGS} is expected to be installed in the system before applying any command listed in Code chunk CC.5.  

\vspace{5pt}
\noindent \textbf{Code chunk CC.5} 
\vspace{-5pt}
\begin{lstlisting}[numbers=left]
R> # Intalling via GitHub:
R> devtools::install_github("vdinizm/pexm")

R> # Installing via Sorceforge: 
R> # Download the appropriate tarball file (Unix or Windows) 
R> # and then replace the argument "local_path_to_source" below by 
R> # the corresponding path to the "tar.gz" file in the computer.  
R> install.packages("local_path_to_source", repos = NULL, type = "source")
\end{lstlisting}

Assuming that \texttt{pexm} was successfully installed in \texttt{R}, the user must first load the PE module before running the MCMC in \texttt{JAGS}. This task is done by typing \texttt{pexm::loadpexm()} in the \texttt{R} console. Further details about this function can be found in the \texttt{pexm} documentation.  

\vspace{5pt}
%%%%%%%%%%%%%%%%%%%%%%%%%%%%%%%%%%%%%%%%%%%%%%%%%%%%%%%%%%%
\renewcommand{\thefigure}{B.\arabic{figure}} \setcounter{figure}{0}
\renewcommand{\theequation}{B.\arabic{equation}} \setcounter{equation}{0}
\renewcommand{\thetable}{B.\arabic{table}} \setcounter{table}{0}
{\flushleft \large \textbf{Appendix B: Additional information for the real application.} }
%%%%%%%%%%%%%%%%%%%%%%%%%%%%%%%%%%%%%%%%%%%%%%%%%%%%%%%%%%%
\vspace{5pt}

The Code chunk CC.6 shows the \texttt{JAGS} script for the zeros-trick implementation of the Bayesian models discussed in Section \ref{secresult}. The prior specification for the rates $\lambda_j$'s is supposed to be included in last line; consider the Code chunk CC.4 (b) for Model I and CC.4 (c) for Model II. Line 2 specifies the time grid, the object \texttt{d[i, k, j]} indicates whether the event-time belongs to the interval $I_j$ and the object \texttt{le[i, k, j]} receives the length of the intersection between the intervals ($0, t_{ik}$) and $I_j$.

\vspace{5pt}
\noindent \textbf{Code chunk CC.6} 
\vspace{-5pt}
\begin{lstlisting}[numbers=left]
data{ 
  for(j in 1:(m+1)) { a[j] <- 562 * (j - 1) / m }
  for(i in 1:n){ for(k in 1:2){ zeros[i, k] <- 0 } }
}
model{ 
  for(i in 1:n){ 
    for(k in 1:2){
      for(j in 1:m){
        d[i, k, j] <- delta[i, k] * step(t[i, k] - a[j]) * step(a[j + 1] - t[i, k])
        le[i, k, j] <- (min(t[i, k], a[j + 1]) - a[j]) * step(t[i, k] - a[j])
        theta[i, k, j] <- lambda[j] * exp( beta_sex * sex[i] + beta_age * age[i,k] ) * z[i]
        mu[i, k, j] <- le[i, k, j] * theta[i, k, j]
        lik[i, k, j] <- dpois(d[i, k, j], mu[i, k, j])
      }
      loglik[i, k] <- sum(log(lik[i, k, ]))
      zeros[i, k] ~ dpois( -loglik[i, k])
    }
  } 
  for(i in 1:n){ z[i] ~ dgamma(eta, eta) }
  kappa <- 1 / eta
  eta ~ dgamma(0.001, 0.001)
  beta_sex ~ dnorm(0, 0.001)
  beta_age ~ dnorm(0, 0.001)
  ... # Include here the Code chunk CC.4 (b) or (c).
}
\end{lstlisting}

%%%%%%%%%%%%%%%%%%%%%%%%%%%%
%%%%%%%%%%%%%%%%%%%%%%%%%%%%
%% Bibliography BibTeX.
{\bibliographystyle{Chicago}
\setlength{\bibsep}{0.3pt}
\bibliography{references}}

\begin{thebibliography}{}

\bibitem[\protect\citeauthoryear{Aslanidou, Dey, and Sinha}{Aslanidou
  et~al.}{1998}]{Asl98}
Aslanidou, H., D.~K. Dey, and D.~Sinha (1998).
\newblock Bayesian analysis of multivariate survival data using monte carlo
  methods.
\newblock {\em Canadian Journal of Statistics\/}~{\em 26}, 33--48.

\bibitem[\protect\citeauthoryear{Barbosa, Colosimo, and Neto}{Barbosa
  et~al.}{1996}]{Bar96}
Barbosa, E.~P., E.~A. Colosimo, and F.~L. Neto (1996).
\newblock Accelerated life tests analyzed by a piecewise exponential
  distribution via generalized linear models.
\newblock {\em IEEE Transactions on Reliability\/}~{\em 45\/}(4), 619--623.

\bibitem[\protect\citeauthoryear{Breslow}{Breslow}{1972}]{Bres72}
Breslow, N. (1972).
\newblock Discussion on regression models and life-tables (by d. r. cox).
\newblock {\em Journal of the Royal Statistical Society B\/}~{\em 34},
  216--217.

\bibitem[\protect\citeauthoryear{Breslow}{Breslow}{1974}]{Bres74}
Breslow, N. (1974).
\newblock Covariance analysis of censored survival data.
\newblock {\em Biometrics\/}~{\em 30}, 89--99.

\bibitem[\protect\citeauthoryear{Brezger, Kneib, and Lang}{Brezger
  et~al.}{2005}]{Brez05}
Brezger, A., T.~Kneib, and S.~Lang (2005).
\newblock Bayesx: Analyzing bayesian structured additive regression models.
\newblock {\em Journal of Statistical Software\/}~{\em 14\/}(11), 1--22.

\bibitem[\protect\citeauthoryear{Clark and Ryan}{Clark and Ryan}{2002}]{Cla02}
Clark, D.~E. and L.~M. Ryan (2002).
\newblock Concurrent prediction of hospital mortality and length of stay from
  risk factors on admission.
\newblock {\em Health Services Research\/}~{\em 37}, 631--645.

\bibitem[\protect\citeauthoryear{Clayton and Cuzick}{Clayton and
  Cuzick}{1985}]{Cla85}
Clayton, D.~G. and J.~Cuzick (1985).
\newblock Multivariate generalizations of the proportional hazards model (with
  discussion).
\newblock {\em Journal of the Royal Statistical Society A\/}~{\em 148},
  82--117.

\bibitem[\protect\citeauthoryear{Cox}{Cox}{1972}]{Cox72}
Cox, D.~R. (1972).
\newblock Regression models and life-tables (with discussion).
\newblock {\em Journal of the Royal Statistical Society B\/}~{\em 34},
  187--220.

\bibitem[\protect\citeauthoryear{Demarqui, Dey, Loschi, and Colosimo}{Demarqui
  et~al.}{2014}]{Dem14}
Demarqui, F.~N., D.~K. Dey, R.~H. Loschi, and E.~A. Colosimo (2014).
\newblock Fully semiparametric bayesian approach for modeling survival data
  with cure fraction.
\newblock {\em Biometrical Journal\/}~{\em 56\/}(2), 198--218.

\bibitem[\protect\citeauthoryear{Denwood}{Denwood}{2016}]{Den16}
Denwood, M.~J. (2016).
\newblock runjags: An r package providing interface utilities, model templates,
  parallel computing methods and additional distributions for mcmc models in
  jags.
\newblock {\em Journal of Statistical Software\/}~{\em 71\/}(9), 1--25.

\bibitem[\protect\citeauthoryear{Gamerman}{Gamerman}{1991}]{Gam91}
Gamerman, D. (1991).
\newblock Dynamic bayesian models for survival data.
\newblock {\em Applied Statistics\/}~{\em 40\/}(1), 63--79.

\bibitem[\protect\citeauthoryear{Gamerman}{Gamerman}{1994}]{Gam94}
Gamerman, D. (1994).
\newblock Bayes estimation of the piecewise exponential distribution.
\newblock {\em IEEE Transactions on Reliability\/}~{\em 43}, 128--131.

\bibitem[\protect\citeauthoryear{Gamerman and Lopes}{Gamerman and
  Lopes}{2006}]{Gam06}
Gamerman, D. and H.~F. Lopes (2006).
\newblock {\em Markov Chain Monte Carlo: Stochastic Simulation for Bayesian
  Inference\/} (2 ed.), Volume~68.
\newblock London: Chapman and Hall/CRC.

\bibitem[\protect\citeauthoryear{Gelfand and Smith}{Gelfand and
  Smith}{1990}]{Gelf90}
Gelfand, A.~E. and A.~F.~M. Smith (1990).
\newblock Sampling-based approaches to calculating marginal densities.
\newblock {\em Journal of the American Statistical Association\/}~{\em
  85\/}(410), 398--409.

\bibitem[\protect\citeauthoryear{Geman and Geman}{Geman and
  Geman}{1984}]{Gem84}
Geman, S. and D.~Geman (1984).
\newblock Stochastic relaxation, gibbs distributions, and the bayesian
  restoration of images.
\newblock {\em IEEE Transactions on Pattern Analysis and Machine
  Intelligence\/}~{\em 6}, 721--741.

\bibitem[\protect\citeauthoryear{Gilks}{Gilks}{}]{Gil92c}
Gilks, W.
\newblock {\em Bayesian Statistics 4}, Chapter {Derivative-Free Adaptive
  Rejection Sampling for Gibbs Sampling}, Pages = {641-649}, Publisher =
  {Oxford University Press}, Year = {1992}.

\bibitem[\protect\citeauthoryear{Gilks, Best, and Tan}{Gilks
  et~al.}{1995}]{Gil95}
Gilks, W., N.~Best, and K.~Tan (1995).
\newblock Adaptive rejection metropolis sampling within gibbs sampling.
\newblock {\em Journal of Royal Statistical Society, Series C\/}~{\em 44},
  455--472.

\bibitem[\protect\citeauthoryear{Gilks and Wild}{Gilks and Wild}{1992}]{Gil92}
Gilks, W. and P.~Wild (1992).
\newblock Adaptive rejection sampling for gibbs sampling.
\newblock {\em Journal of the Royal Statistical Society, Series C\/}~{\em 41},
  337--348.

\bibitem[\protect\citeauthoryear{Hastings}{Hastings}{1970}]{Has70}
Hastings, W. (1970).
\newblock Monte carlo sampling using markov chains and their applications.
\newblock {\em Biometrika\/}~{\em 57}, 97--109.

\bibitem[\protect\citeauthoryear{Hoffman and Gelman}{Hoffman and
  Gelman}{2014}]{Hoff14}
Hoffman, M.~D. and A.~Gelman (2014).
\newblock The no-u-turn sampler: Adaptively setting path lengths in hamiltonian
  monte carlo.
\newblock {\em Journal of Machine Learning Research\/}~{\em 15}, 1351--1381.

\bibitem[\protect\citeauthoryear{Ibrahim, Chen, and Sinha}{Ibrahim
  et~al.}{2001}]{Ibra01}
Ibrahim, J.~G., M.~H. Chen, and D.~Sinha (2001).
\newblock {\em Bayesian Survival Analysis}.
\newblock Springer series in statistics. New York: Springer-Verlag.

\bibitem[\protect\citeauthoryear{Jackson}{Jackson}{2011}]{Jack11}
Jackson, C.~H. (2011).
\newblock Multi-state models for panel data: The msm package for r.
\newblock {\em Journal of Statistical Software\/}~{\em 38\/}(8), 1--29.

\bibitem[\protect\citeauthoryear{Kalbfleisch and Prentice}{Kalbfleisch and
  Prentice}{1973}]{Kalb73}
Kalbfleisch, J.~D. and R.~L. Prentice (1973).
\newblock Marginal likelihoods based on cox's regression and life model.
\newblock {\em Biometrika\/}~{\em 60}, 267--278.

\bibitem[\protect\citeauthoryear{Kim and Proschan}{Kim and
  Proschan}{1991}]{Kim91}
Kim, J.~A. and F.~Proschan (1991).
\newblock Piecewise exponential estimator of the survivor function.
\newblock {\em IEEE Transactions on Reliability\/}~{\em 40}, 134--139.

\bibitem[\protect\citeauthoryear{Lunn, Jackson, Best, Thomas, and
  Spiegelhalter}{Lunn et~al.}{2012}]{Lun12}
Lunn, D., C.~Jackson, N.~G. Best, A.~Thomas, and D.~J. Spiegelhalter (2012).
\newblock {\em The BUGS Book: A Practical Introduction to Bayesian Analysis}.
\newblock Boca Raton: Chapman and Hall/CRC.

\bibitem[\protect\citeauthoryear{McGilchrist and Aisbett}{McGilchrist and
  Aisbett}{1991}]{Mcgil91}
McGilchrist, C.~A. and C.~W. Aisbett (1991).
\newblock Regression with frailty in survival analysis.
\newblock {\em Biometrics\/}~{\em 47}, 461--466.

\bibitem[\protect\citeauthoryear{Metropolis, Rosenbluth, Teller, and
  Teller}{Metropolis et~al.}{1953}]{Met53}
Metropolis, N., A.~Rosenbluth, M.~Teller, and E.~Teller (1953).
\newblock Equations of state calculations by fast computing machines.
\newblock {\em Journal of Chemistry and Physics\/}~{\em 21}, 1087--1091.

\bibitem[\protect\citeauthoryear{Neal}{Neal}{2003}]{Neal03}
Neal, R. (2003).
\newblock Slice sampling.
\newblock {\em The Annals of Statistics\/}~{\em 31}, 705--767.

\bibitem[\protect\citeauthoryear{Oakes}{Oakes}{1986}]{Oak86}
Oakes, D. (1986).
\newblock Semiparametric inference in a model for association in bivariate
  survival data.
\newblock {\em Biometrika\/}~{\em 73}, 353--361.

\bibitem[\protect\citeauthoryear{Oakes}{Oakes}{1989}]{Oak89}
Oakes, D. (1989).
\newblock Bivariate survival models induced by frailties.
\newblock {\em Journal of the American Statistical Association\/}~{\em 84},
  487--493.

\bibitem[\protect\citeauthoryear{Plummer}{Plummer}{2003}]{Plu03}
Plummer, M. (2003).
\newblock Jags: A program for analysis of bayesian graphical models using gibbs
  sampling.

\bibitem[\protect\citeauthoryear{Plummer}{Plummer}{2010}]{Plu10}
Plummer, M. (2010).
\newblock {\em JAGS Developers Manual}.

\bibitem[\protect\citeauthoryear{Plummer}{Plummer}{2016}]{Plu16}
Plummer, M. (2016).
\newblock {\em rjags: Bayesian Graphical Models Using MCMC}.
\newblock R package version 4-5.

\bibitem[\protect\citeauthoryear{Plummer}{Plummer}{2017}]{Plu17}
Plummer, M. (2017).
\newblock {\em JAGS Version 4.3.0 User Manual}.

\bibitem[\protect\citeauthoryear{Plummer, Best, Cowles, and Vines}{Plummer
  et~al.}{2006}]{Plu06}
Plummer, M., N.~Best, K.~Cowles, and K.~Vines (2006).
\newblock Coda: Convergence diagnosis and output analysis for mcmc.
\newblock {\em R News\/}~{\em 6\/}(1), 7--11.

\bibitem[\protect\citeauthoryear{Rigdon and Basu}{Rigdon and
  Basu}{2000}]{Rig00}
Rigdon, S.~E. and A.~P. Basu (2000).
\newblock {\em Statistical Methods for the Reliability of Repairable Systems}.
\newblock New York: John Wiley and Sons.

\bibitem[\protect\citeauthoryear{Rue, Martino, and Chopin}{Rue
  et~al.}{2009}]{Rue09}
Rue, H., S.~Martino, and N.~Chopin (2009).
\newblock Approximate bayesian inference for latent gaussian models by using
  integrated nested laplace approximations.
\newblock {\em Journal of the Royal Statistical Society, Series B\/}~{\em
  71\/}(2), 319--392.

\bibitem[\protect\citeauthoryear{Sahu, Dey, Aslanidou, and Sinha}{Sahu
  et~al.}{1997}]{Sahu97}
Sahu, S.~K., D.~K. Dey, H.~Aslanidou, and D.~Sinha (1997).
\newblock A weibull regression model with gamma frailties for multivariate
  survival data.
\newblock {\em Lifetime Data Analysis\/}~{\em 3}, 123--137.

\bibitem[\protect\citeauthoryear{Sinha, Chen, and Ghosh}{Sinha
  et~al.}{1999}]{Sinha99}
Sinha, D., M.~H. Chen, and S.~K. Ghosh (1999).
\newblock Bayesian analysis and model selection for interval-censored survival
  data.
\newblock {\em Biometrics\/}~{\em 55}, 585--590.

\bibitem[\protect\citeauthoryear{Sinha and Dey}{Sinha and Dey}{1997}]{Sinha97}
Sinha, D. and D.~K. Dey (1997).
\newblock Semiparametric bayesian analysis of survival data.
\newblock {\em Journal of the American Statistical Association\/}~{\em
  92\/}(439), 1195--1212.

\bibitem[\protect\citeauthoryear{Spiegelhalter, Thomas, and Best}{Spiegelhalter
  et~al.}{2003}]{Spie03}
Spiegelhalter, D.~J., A.~Thomas, and N.~G. Best (2003).
\newblock {\em WinBUGS Version 1.4 User Manual}.
\newblock Cambridge: Medical Research Council Biostatistics Unit.

\bibitem[\protect\citeauthoryear{Spiegelhalter, Thomas, Best, and
  Lunn}{Spiegelhalter et~al.}{2014}]{Spie14}
Spiegelhalter, D.~J., A.~Thomas, N.~G. Best, and D.~Lunn (2014).
\newblock {\em OpenBUGS User Manual}.
\newblock Version 3.2.3.

\bibitem[\protect\citeauthoryear{{Stan Development Team}}{{Stan Development
  Team}}{2019}]{Rstan}
{Stan Development Team} (2019).
\newblock rstan: the r interface to stan.
\newblock R package version 2.19.2.

\bibitem[\protect\citeauthoryear{Team}{Team}{2020}]{R}
Team, R.~C. (2020).
\newblock {\em R: A Language and Environment for Statistical Computing}.
\newblock Vienna, Austria: R Foundation for Statistical Computing.

\bibitem[\protect\citeauthoryear{Therneau}{Therneau}{2015}]{The15}
Therneau, T. (2015).
\newblock {\em A Package for Survival Analysis in S}.
\newblock R package version 2.38.

\bibitem[\protect\citeauthoryear{Thomas, Spiegelhalter, and Gilks}{Thomas
  et~al.}{1992}]{Tho92}
Thomas, A., D.~Spiegelhalter, and W.~Gilks (1992).
\newblock Bugs: A program to perform bayesian inference using gibbs sampling.
\newblock {\em Bayesian Statistics\/}~{\em 4}, 837--942.

\bibitem[\protect\citeauthoryear{Vaupel, Manton, and Stallard}{Vaupel
  et~al.}{1979}]{Vau79}
Vaupel, J.~M., K.~G. Manton, and E.~Stallard (1979).
\newblock The impact of heterogeneity in individual frailty on the dynamics of
  mortality.
\newblock {\em Demography\/}~{\em 16}, 439--454.

\bibitem[\protect\citeauthoryear{Wabersich and Vandekerckhove}{Wabersich and
  Vandekerckhove}{2014}]{Wab14}
Wabersich, D. and J.~Vandekerckhove (2014).
\newblock Extending jags: A tutorial on adding custom distributions to jags
  (with a diffusion model example).
\newblock {\em Behavior Research Methods\/}~{\em 46}, 15--28.

\bibitem[\protect\citeauthoryear{Wickham, Hester, and Chang}{Wickham
  et~al.}{2020}]{devtools}
Wickham, H., J.~Hester, and W.~Chang (2020).
\newblock {\em devtools: Tools to Make Developing R Packages Easier}.
\newblock R package version 2.2.2.

\end{thebibliography}

%%%%%%%%%%%%%%%%%%%%
\end{document}